\documentclass[]{aastex631}

\shorttitle{Comparative Mid-Infrared Spectroscopy of Dark, Primitive Asteroids}
\shortauthors{Humes et al.}

\graphicspath{{./}{figures/}}

\begin{document}

\title{Comparative Mid-Infrared Spectroscopy of Dark, Primitive Asteroids: Does Shared Taxonomic Class Indicate Shared Silicate Composition?}

\correspondingauthor{Oriel Humes}
\email{oah28@nau.edu}

\author[0000-0002-1700-5364]{Oriel A. Humes}
\affiliation{Northern Arizona University \\
Flagstaff, Arizona \\
86011, USA}

\author[0000-0003-3402-1339]{Audrey C. Martin}
\affiliation{University of Central Florida \\
Orlando, Florida \\
32816, USA}

\author[0000-0003-3091-5757]{Cristina A. Thomas}
\affiliation{Northern Arizona University \\
Flagstaff, Arizona \\
86011, USA}

\author[0000-0001-9265-9475]{Joshua P. Emery}
\affiliation{Northern Arizona University \\
Flagstaff, Arizona \\
86011, USA}

\begin{abstract}

 Primitive asteroids with low albedos and red slopes in the visible and near infrared (VNIR) are found in both the Main Belt and the Jupiter Trojan clouds. In order to determine whether the VNIR spectral similarities of primitive Main Belt asteroids and Jupiter Trojans are reflective of a true compositional similarity, we compare the mid-infrared silicate emission features of Main Belt and Jupiter Trojan asteroids. Using archival data from the Spitzer Space Telescope's IRS spectrograph and observations from the Stratospheric Observatory for Infrared Astronomy's (SOFIA) FORCAST instrument, we analyze the 5-40 $\mu$m spectra of thirteen primitive Main Belt asteroids and compare them to those of Jupiter Trojans in the literature. We find that while many primitive asteroids in the Main Belt resemble their Trojan counterparts with strong spectral signatures of olivine-rich high-porosity silicate regoliths, we identify (368) Haidea as a spectrally distinctive asteroid that lacks strong evidence of olivine in its MIR spectrum.  Differences in silicate compositions among D-type asteroids imply a diversity of origins for primitive asteroids. 

\end{abstract}


\section{Introduction} \label{sec:intro}

Small Solar System bodies, including asteroids and comets, are understood as tracers of the evolutionary history of the Solar System. Primitive asteroids---those that have undergone little to no geothermal evolution since their formation---in particular are valuable sources of information about conditions in the early Solar System as their compositions closely reflect the composition of the Solar nebula in which they formed. The present-day compositional distribution of asteroids is thought to reflect a combination of initial conditions in the protosolar nebula and the subsequent transport of small bodies (\cite{demeo2014}). In general for large ($>$50 km) asteroids, the abundance of high-melting point materials at low heliocentric distances and abundance of relatively more volatile materials at high heliocentric distances suggests that these compositions reflect thermal and chemical gradients within the early Solar System (\cite{gradie1982, demeo2014}). For instance, the presence of hydrated silicates in carbonaceous chondrites, analogs for C-type asteroids common in the Main Belt, suggests a formation region beyond the water frost line but not so far from the Sun that long orbital timescales preclude the accretion of asteroids prior to the decay of short-lived radionuclides that provide the heat source necessary for thermal alteration to occur (\cite{grimm1989}). Thus, the locations of primitive C-type asteroids can indicate the availability of water within the solar nebula (e.g. Ceres in \cite{mccord2005}). On the other hand, the diversity of all asteroid types at small scales throughout the Main Belt (\cite{demeo2014_dtype, demeo2014}) provides evidence for substantial migration and mixing among small bodies after their formation. Both initial compositional gradients and subsequent mixing of asteroid materials are necessary to explain the distribution of asteroids we see today.

The D- \& P-type asteroids, characterized by their low albedos and red sloped, featureless spectra (\cite{tholen1984asteroid, tholen1989asteroid, DeMeo2009}) in the visible and near infrared (VNIR), are a particularly intriguing population of primitive asteroids (\cite{VILAS1985}). Dynamical modeling has led to the suggestion that Trojan (\cite{Morbidelli_2005}) and Main Belt (\cite{levison2009}) D- and P-types originated from the relatively cooler outer regions of the protoplanetary disk. Their increasing frequency at larger heliocentric distances (\cite{tholen1989asteroid, demeo2014}) and spectral resemblance to some outer Solar System populations including Centaurs and trans-Neptunian Objects (\cite{sheppard2010colors}) supports this idea. As primitive asteroids, the D- \& P-types are thought to have undergone little to no internal differentiation and hydrothermal alteration, thus preserving much of their original compositions (\cite{jones1990composition}). If the D- \& P-types preserve their original compositions, understanding their compositions and distribution may provide unique insights into chemical and dynamical conditions in the early Solar System. Previous spectral modeling studies of D- \& P-types in the VNIR (\cite{emery2004, sharkey2019, gartrelle2021illuminating}) have proposed compositions rich in silicates with trace amounts of organics and ices, suggestive of an outer Solar System origin. However, the lack of VNIR spectral features limits our ability to infer the compositions of these asteroids.  Understanding the origin of primitive asteroids is a priority for a number of recent and upcoming spacecraft missions including the Lucy (\cite{Levison_2021}) mission.

The Jupiter Trojan population is a major reservoir for D- \& P-type asteroids, which comprise the majority of asteroids in the Jovian Trojan swarms (\cite{emery2010near, demeo2014}). The Jupiter Trojans orbit in proximity to the Sun-Jupiter L4 and L5 Lagrange points, and because of their unique orbital relationship to Jupiter, many dynamical hypotheses have been proposed to explain their emplacement within the Lagrange regions (e.g. \cite{Morbidelli_2005, Nesvorny_2013, pirani_2019}). These models demonstrate that the properties of the Trojans we observe today depend both on where the Trojans initially formed and the subsequent migration history of the giant planets that led to their emplacement within Jupiter's Lagrange regions.  For instance, the Nice model (\cite{Morbidelli_2005}) predicts a primarily outer Solar System origin for Trojans, while the Jumping Jupiter (\cite{Nesvorny_2013}) model predicts that the Trojans sample multiple regions including both the inner and outer Solar System. Probing the relationship between D- \& P-types in the Main Belt and Trojan swarms can provide an observational test for these models of giant planet migration. For example, the Nice model predicts a common, specifically outer Solar System, parent population for D- \& P-types across the Main Belt, Hilda, and Trojan populations (\cite{levison2009contamination}).

In contrast to the relatively featureless VNIR, in the mid-infrared, many D- \& P-type asteroids exhibit 10- and 20-$\mu$m silicate emission features (\cite{emery_2006, mueller2010, licandro_2011}). These features are diagnostic of both physical and chemical properties of silicates, including regolith porosity (\cite{martin2022_olivine, MARTIN_pyroxene}), Mg\# (the ratio of magnesium to iron in a mineral lattice) for both pyroxenes (\cite{hamilton2000pyroxene, chihara2002pyroxene}) and olivines (\cite{koike2003olivine, hamilton2010olivine, lane2011olivine}), amorphous vs crystalline mineral structure (\cite{martin-trojan}), and silicate hydration state (\cite{Beck2014,McAdam2015}. The compositional properties as revealed by silicate emission features in the mid-infrared ($\sim$ 5-40 $\mu$m) are indicative of formation region, as temperature conditions within the initial solar nebula influence the materials and formation pathways available to condense silicates at varying heliocentric distances (\cite{gail2004,henning2010cosmic,morlok2014}). In \cite{martin-trojan} the silicate emissivity spectrum was characterized in detail to determine composition and regolith porosity of several Jupiter Trojans. That study concluded that Jupiter Trojans likely originate from the outer Solar System due to the abundance of magnesium-rich silicates, absence of hydrated silicates, and high porosity regoliths similar to those of comets.

\begin{deluxetable*}{lccccc}[ht]
\tablenum{1}
\tablecaption{Physical and orbital properties of primitive asteroids examined in this paper \label{tab:asteroid_list}}
\tablewidth{0pt}
\tablehead{
 & \colhead{Semimajor} Axis & & \colhead{Inclination} & \colhead{H} &    \\
\colhead{Asteroid} & \colhead{(AU)} & \colhead{Eccentricity} & \colhead{(deg)} & \colhead{(mag.)} & \colhead{Taxonomic Classification} 
  }
\startdata
46 Hestia	         &2.52	&0.17	&2.35	&8.59	&P/Xc    \\
65 Cybele	         &3.44	&0.12	&3.56	&6.89	&P/Xc      \\        
87 Sylvia	         &3.48	&0.09	&10.9	&6.98	&P/X      \\
140 Siwa	         &2.73	&0.21	&3.19	&8.5	&P/Xc     \\
190 Ismene	         &3.99	&0.17	&6.18	&7.7	&P/X      \\
267 Tirza 	         &2.78	&0.10	&6.01	&10.3	&DU/D     \\
269 Justitia	     &2.62	&0.21	&5.48	&9.79	&Ld/IR-RR \\
336 Lacadiera	     &2.25	&0.10	&5.65	&9.86	&D/Xk    \\
368 Haidea 	         &3.07	&0.20	&7.8	&10	    &D/       \\
944 Hidalgo	         &5.73	&0.66	&42.6	&10.56	&D/       \\
1268 Libya	         &3.98	&0.10	&4.43	&9.12	&P/       \\
1702 Kalahari	     &2.86	&0.14	&9.96	&11.1	&D/L      \\
3552 Don Quixote	 &4.26	&0.71	&31.1	&13.07	&D/D     \\
\enddata
\tablecomments{Taxonomic classifications are given as (Tholen/SMASSII) classifications from \cite{PDS_tax}, except in the cases of Justitia, which has been classified using Tholen and the TNO classification scheme from \cite{hasegawa2021} and Don Quixote, which uses the Tholen/SMASSII classification and comes from from \cite{binzel2004}. Note that the IR-RR classification for Justitia indicates that it is redder than the typical D type. }
\end{deluxetable*}

In this paper we examine the spectra of thirteen D- \& P-type asteroids as observed by the Stratospheric Observatory for Infrared Astronomy (SOFIA) and Spitzer Space Telescope (see Table \ref{tab:asteroid_list}). This group samples objects in the Main Belt, Cybele, and near-Earth populations, but not the Trojans previously characterized in \cite{martin-trojan}. We examine silicate emission features in the mid-infrared to compare the spectra of Main Belt D- \& P-types and literature spectra of Jupiter Trojan D- \& P-types (\cite{martin-trojan, emery_2006}) to determine whether these two populations, which are similar in the VNIR, resemble each other in the mid-infrared. Similarities in silicate compositions among these two populations would point to a shared region of origin for primitive asteroids across the Trojan and Main Belt populations, while compositional differences would indicate separate parent populations for primitive Main Belt and Trojan asteroids. Determining whether primitive asteroids in the Main Belt share the outer Solar System origin suggested by \cite{martin-trojan} for the Trojans will allow us to determine the extent of mixing between inner and outer Solar System materials within the Main Belt and provide new observational constraints on formation and emplacement mechanisms for primitive asteroids. Constraining the amount of outer Solar System material that has been delivered to the Main Belt and inner Solar System subsequently will lead to a greater understanding of the mechanisms responsible for delivering volatiles from beyond the frost line to the inner Solar System.

\section{Methods}
\subsection{Spitzer Space Telescope Observations}

We identified eleven non-Trojan D- \& P-type asteroids in the Spitzer Heritage archive. These asteroids were observed using the Infrared Spectrograph (IRS) instrument (\cite{houck2004IRS}) by programs IRS\_P/666 and IRS\_R/668 (IRS Campaigns P \& R by PI James R Houck);  and ASTEROIDS-1/88 and ASTEROIDS-2/91 (Extinct Comets and Low-Albedo Asteroids \& Extinct Comets and Low-Albedo Asteroids 2 by PI Dale Cruikshank). These targets included observations in all four IRS modes: short-low (SL, 5.2-14.5 $\mu$m), long-low (LL, 14.0-38.0 $\mu$m), short-high (SH, 9.9-19.6 $\mu$m), and long-high (LH, 18.7-37.2$\mu$m). Observations in any one of these modes are useful for identifying silicate emission features, as the SL and SH modes cover the 10 $\mu$m region and the LL and LH modes cover the 20 $\mu$m region, and the choice of observing mode was influenced by the brightness of individual targets. For observations in the low resolution modes (SL and LL), background pair subtraction was performed by pairing images in two different nod positions. In some cases (e.g. Siwa), the two nod positions resulted in the spectral trace falling on two different regions of the same section of the chip, while most asteroids had spectra that nodded between the two different sections of the chip (i.e. orders 1 \& 2 vs. order 3). For observations in the high resolution modes (SH and LH),  background subtraction was done by subtracting the average of “shadow” observations taken of empty sky at the observed location of the asteroid after some time had passed and the asteroid had moved. While SH and LH spectra were available for Siwa and Lacadiera in these modes, the lack of corresponding “shadow” observations prevented these additional spectra from being extracted since they lacked an appropriate measurement of background.

\begin{deluxetable*}{llllcccc}[ht]
\tablenum{2}
\tablecaption{Observation circumstances for D- \& P-type asteroids in the Spitzer Heritage Archive \label{tab:spitzer_obs}}
\tablewidth{0pt}
\tablehead{
 & & & \colhead{Obs. Time}  & \colhead{Exp. Time} & \colhead{$r$} & \colhead{$\Delta$} & \colhead{Phase Angle}   \\
\colhead{Asteroid} & \colhead{Modes} & \colhead{Program} & \colhead{UTC} & \colhead{(sec)} & \colhead{AU} & \colhead{AU} & \colhead{deg} 
  }
\startdata
65 Cybele   &	Spitzer SH	& ASTEROIDS-2/91 &	2005-Feb-08 6:05 &	   12      &3.79 &	3.30 &	14.1   \\
            &	Spitzer SL &	ASTEROIDS-2/91	& 2005-Feb-08 6:00 &	43     & 3.79 &	3.3	& 14.1 \\
87 Sylvia   &	Spitzer SH, LH	& ASTEROIDS-1/88 &	2004-Nov-12 6:40 &	12, 13    & 3.22 &	2.71 &	17.1  \\
	        & Spizter SL &	ASTEROIDS-1/88 &	2004-Nov-12 5:47 &	   26      & 3.22 &	2.71 &	17.1 \\
140 Siwa    &	Spitzer SL &	ASTEROIDS-2/91 &	2004-Feb-29 9:01 &	 14    & 2.65 &	2.23 &	21.4   \\
267 Tirza   &	Spitzer SL, LL	& ASTEROIDS-2/91 &	2005-Nov-17 1:40 &	  135, 25   & 3.02 &	2.97 &	19.6   \\
269 Justitia & Spitzer SL & ASTEROIDS-2/91 & 2005-Apr-15 7:54 &      25     & 2.12	& 1.66	& 27.5 \\
336 Lacadiera & Spitzer SL &	ASTEROIDS-2/91 &	2004-Oct-03 17:29 &	  24   & 2.43 &	1.96 &	24.0 \\
368 Haidea   &	Spitzer SL, LL &	ASTEROIDS-1/88 &	2004-Mar-25 23:29 &	46, 25 & 3.69 &	3.15 &	14.1 \\
944 Hidalgo  &	Spitzer SL, LL &	ASTEROIDS-2/91 &	2006-Jul-24 10:18 &	53, 25 &1.96 &	1.71 &	30.7   \\
1268 Libya    &	Spitzer SL, LL &	IRS\_P/666, IRS\_R/668 &	2003-Nov-16 10:32 &	     609, 148       & 4.19  & 4.00&	13.9   \\
1702 Kalahari &	Spitzer SL, LL &	ASTEROIDS-2/91 &	2005-Jul-03 4:42 &	513, 58  & 3.09 &	2.67 &	18.6  \\
3552 Don Quixote &	Spitzer LL &	ASTEROIDS-1/88 &	2004-Mar-23 4:43 &	4696  & 6.91 &	6.49 &	7.76  \\
\enddata
\tablecomments{For each asteroid, we report the modes used to observe the asteroid, the associated observing program, the mean observation time, total exposure time for each mode, as well as the observational geometries including heliocentric distance $r$, distance from the asteroid to the telescope $\Delta$, and the phase angle as calculated by the JPL Horizons online ephemeris.  }
\end{deluxetable*}

After background subtraction, the spectra were extracted using the Spitzer IRS Custom Extraction software (SPICE), following the general procedure outlined in Recipe 16 of the Spitzer Data Analysis Cookbook (\cite{SPICE_cookbook}). Next, the extracted spectra were adjusted so that each order was scaled to the same flux value in regions where the orders overlapped. This was accomplished by fitting low (first, second, or third) order polynomials to the data, then determining the resulting multiplicative scaling factor necessary such that the polynomials would have the same value--that is the mean of the predicted fluxes--at a specific wavelength. For some data, obvious outliers due to instrumental effects (e.g. the SL teardrop at long wavelengths) were excluded from the fit to ensure better alignment of the orders. For the SL and LL observations, the alignment wavelengths were 7.5 and 21 $\mu$m respectively, that is, each spectrum was scaled to the average predicted flux measurement of all orders at 7.5 or 21 $\mu$m. For asteroids with both SL and LL observations, we aligned the two modes to the average predicted flux at 15 $\mu$m. For the SH and LH observations, no single wavelength region overlaps in all spectral orders, so predicted fluxes and scale factors required to align the spectral orders were calculated at wavelengths of 10.5, 11, 11.65, 12.4, 13.1, 14.3, 15.1, 16.4, and 17.8 $\mu$m for the SH and 20.5, 21, 22.5, 24, 25.5, 27, 29, 31.25, and 34 $\mu$m  for the LH data, then scaled by the average of the all scale factors calculated for each mode. This has the effect of weighting each region of overlap equally. This scaling process is similar to the one described in \cite{emery_2006}.

For most asteroids, only two spectra were taken, but for some especially bright targets (e.g. the SL observations of Cybele, Sylvia, Justitia, Lacadiera, Hidalgo, Libya, Tirza, and Siwa), IRS was used in mapping mode to ensure the target was well centered in the slit. Rather than using the peak-up cameras to locate the centroid of the target, which saturated for bright targets, observations were taken by stepping across the disk of the asteroid to ensure at least one pair of observations were well centered in the slit. In these instances, only the brightest pair of extracted spectra were used in the rest of the analysis to exclude observations that were not well-centered in the slit. While LH observations of Cybele and corresponding shadow observations exist in the Spitzer Heritage Archive, the fluxes at the long wavelength end of the SH data did not match the fluxes at the short wavelength end of the LH data at 19 $\mu$m, indicating that in the LH data the asteroid was not well centered in the slit. Therefore, we exclude the LH data for Cybele in our analysis. A summary of the Spitzer observations and their observing circumstances can be found in Table \ref{tab:spitzer_obs}.

We note that the Spitzer IRS spectra of several asteroids already exist in the published literature. In particular, Spitzer IRS spectra of 65 Cybele (\cite{licandro_2011}), 87 Sylvia (\cite{marchis2012sylvia}), 386 Haidea (\cite{vernazza2013}), 944 Hidalgo (\cite{lowry2022t}) and 3552 Don Quixote (\cite{mommert2014DQ}) have been examined by previous authors. In these cases we re-reduced the data starting from the Level 2 data products available in the Spitzer Heritage Archive in ensure consistency between all asteroids in our dataset in the reduction and thermal modeling process.

\subsection{Stratospheric Observatory For Infrared Astronomy (SOFIA) Observations}

In addition to the Spitzer observations described above, we also observed four primitive asteroids using the Stratospheric Observatory For Infrared Astronomy (SOFIA). We observed these asteroids using the Faint Object infraRed Camera for the SOFIA Telescope (FORCAST) instrument (\cite{herter2018forcast}) as part of program 08\_0105 (Mid-Infrared Spectroscopy of Primitive Asteroids in the Middle Solar System by PI Oriel Humes). Each asteroid was observed using the G111 and G227 grisms along with contemporaneous imagery in the F111 and F253 filters. Initial processing of these images, including background subtraction and spectral extraction were performed by the Universities Space Research Association (USRA) and the extracted spectra and background-subtracted images were downloaded directly from the Infrared Science Archive (IRSA). We located the centroid of each asteroid in the F111 and F253 images then performed background-subtracted aperture photometry using an object radius of 12 pixels and an background annulus with inner and outer radii of 15 and 25 pixels, respectively, according to the procedure described in the \textit{FORCAST Basic Photometry cookbook} (\cite{sofia_photometry}).  We then used the photometric measurements to align the G111 and G227 spectra for each asteroid. By convolving the object spectra and the transmission spectra of the F111 and F253 filters (\cite{sofia_grism}), we computed the predicted F111 and F253 photometric fluxes. Using the ratio between the predicted and actual fluxes, we calculated the multiplicative factor needed to bring the grism spectra into alignment with the measured photometry such that the predicted convolved fluxes were equal to the measured photometry. While this procedure generally follows the procedure outlined in the \textit{FORCAST Grism Spectra: Basic Inspection and Assessment} notebook in \cite{sofia_grism} we used a multiplicative (rather than additive) correcting factor to align our spectra to the photometry (see Figure \ref{fig:NEATM}, right panel). Our rationale is that in order to convert from flux to emissivity, we must divide by a model of the black body emission of the asteroid. An additive correction factor would then distort the shape of the resulting emissivity spectrum, while a multiplicative factor would maintain its shape. Using a multiplicative factor also parallels the process of aligning the spectral orders of the Spitzer data as described above. 

\begin{deluxetable*}{lllcccc}[ht]
\tablenum{3}
\tablecaption{Observation circumstances for primitive asteroids using SOFIA \label{tab:sofia_obs}}
\tablewidth{0pt}
\tablehead{
 & & \colhead{Exposure Times} &  \colhead{Obs.Time}  & \colhead{$r$} & \colhead{$\Delta$} & \colhead{Phase Angle}   \\
\colhead{Asteroid} & \colhead{Filters/Grisms}  & \colhead{(sec)} & \colhead{UTC} &  \colhead{AU} & \colhead{AU} & \colhead{deg} 
  }
\startdata
46 Hestia   &	G227                      &  544                  &	2021-Jun-30 5:32 &	2.63 &	1.87 &	17.5 \\
            &	G111, G227, F111, F253    &  838, 454, 46.4, 36.2 &	2021-Jul-09 5:14 &	2.62 &	1.95 &	19.6 \\
87 Sylvia   &	G111, G227, F111, F253    & 1280, 699, 67.1, 55.8 &	2022-May-24 6:47 &	3.55 &	2.58 &	5.37 \\
140 Siwa    &	G111, G227, F111, F253    & 935, 349, 70.3, 53.7  &	2022-May-27 5:36 &	2.51 &	1.61 &	13.1 \\
190 Ismene  &   G227, F253	              & 1360, 44.6            & 2022-Feb-02 8:33 &	3.41 &	2.46 &	5.17\\
            &	G111, F111	              & 2350, 70.9            & 2022-Feb-09 9:14 &	3.41 &	2.5	& 7.23 \\
\enddata
\tablecomments{For each asteroid, we report the filters used to observe the asteroid, the mean observation time, total exposure time for each filter, and the observational geometries including heliocentric distance $r$, distance from the asteroid to the telescope $\Delta$, and the phase angle as calculated by the JPL Horizons online ephemeris.}
\end{deluxetable*}

\subsection{Thermal Modeling}

\begin{figure}
\gridline{\fig{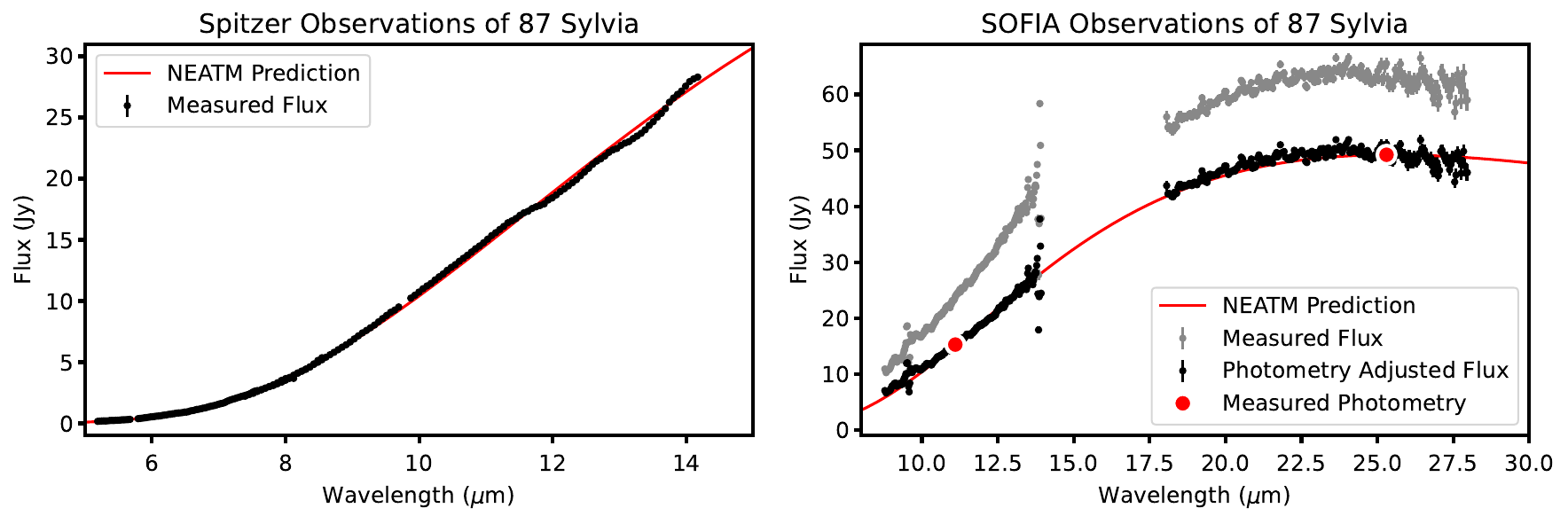}{\textwidth}{}}
\caption{Comparison of the Near Earth Asteroid Thermal Model (NEATM) predictions to the measured fluxes of 87 Sylvia as observed using the Spitzer Space Telescope (left) and the Stratospheric Observatory for Infrared Astronomy (SOFIA, right). For SOFIA observations, spectra are scaled using a multiplicative factor to match the photometric measurements of the asteroid obtained on the same flight (see Section 2.2 of the text for details). The analogous scaling procedure for the Spitzer data is described in Section 2.1 of the text but the unaligned Spitzer data is excluded from the figure as the difference between the aligned and unaligned fluxes is not visible on the scale of the figure. \label{fig:NEATM}}
\end{figure}

Both the Spitzer Space Telescope and SOFIA observe the thermal emission from asteroids in Jansky. In order to convert from a thermal emission spectrum to an emissivity spectrum, we must model the predicted thermal emission for each asteroid, then divide the emission data by this predicted emission spectrum to obtain an emissivity spectrum (\cite{harris_1998}, see Figure \ref{fig:NEATM}). The Near Earth Asteroid Thermal Model (NEATM), while initially developed to model thermal emission of near-Earth asteroids, is commonly used to predict thermal fluxes for Main Belt and Trojan asteroids throughout the literature (e.g. \cite{emery_2006, hargrove_2012, licandro_2011}). By equating incoming solar energy (a factor of the geometric albedo $a$) and outgoing thermal emission, this model computes the thermal emission due to solar heating at each point on the sunlit hemisphere of an asteroid (the contribution from the unlit hemisphere is assumed to be 0). Then, using the geometric circumstances of the observation, the flux received by the observer is computed by integrating the predicted emission over the hemisphere visible to the observer (\cite{harris_1998}). In addition to the geometric factors, the NEATM we used also includes two empirical corrections: a beaming parameter $\eta$ to correct for surface roughness and an emissivity-like multiplicative correction factor $\epsilon$ that represents the ratio of observed to predicted flux. The variation in these empirical correction factors encapsulates the contributions of observational factors not taken into account by the thermal model such as variable surface emissivity and slit losses. 

For our observations, we used the NEATM included in the \texttt{mskpy surfaces} package developed for python by Michael Kelley (\cite{sbpy}). This model uses $\chi^2$ minimization to derive the best fit diameter and beaming parameter $\eta$ given observing geometry, albedo, and the average emissivity and computes the corresponding modeled fluxes at a given wavelength. We assumed an average emissivity of $\epsilon = 0.9$, the mean emissivity for asteroids as reported in \cite{mainzer_2011}. Both diameter and the beaming parameter are free to vary in the $\chi^2$ minimization routine. To convert from flux to emissivity, we divided the fluxes as measured by the telescope by the modeled flux, then multiplied by the assumed average bolometric emissivity. To derive errors on diameter and beaming parameter, we assume an additional uncertainty in flux to account for uncertainty in the absolute photometric calibration of the spectra in addition to reported error measurements. For Spitzer, we assume this uncertainty is 10\%, following the findings of \cite{decin2004marcs}. For the SOFIA data, the absolute flux is calibrated using photometry, and we take the average uncertainty to be (5.6\%), the average error in photometry across all our targets. After computing the NEATM-derived diameter, beaming parameter, and their associated errors, we propagate the expected error due to absolute flux uncertainty by adding the expected uncertainty in quadrature to the NEATM-derived errors. In most cases, the error associated with absolute flux uncertainty dominates the error term. For most asteroids, we used a single NEATM to model the flux, that is to say that both short wavelength spectra (e.g. from Sptizer SL or SOFIA G111) and long wavelength spectra (e.g. from Spitzer LL or SOFIA G227) were combined into a single spectrum that was used as input into the thermal model if the observations were taken on the same day. For 368 Haidea and 1702 Kalahari, we found that the long wavelength end of the Spitzer SL spectrum was heavily affected by the SL teardrop, a noted artifact affecting this region of Spitzer spectra. The artifact led to difficulty in aligning the SL and LL spectra. For these two asteroids, we fit a separate NEATM for the SL and LL segments. 

\begin{deluxetable*}{llcccc}[ht]
\tablenum{4}
\tablecaption{Results of Thermal Modeling \label{tab:NEATM}}
\tablewidth{0pt}
\tablehead{
\colhead{Asteroid} & \colhead{Instrument/Mode} & \colhead{Literature Diameter} &\colhead{Literature } & \colhead{Derived Diameter} & \colhead{Derived Beaming}\\
& & (km) & Albedo  & (km) & Parameter}
\startdata
46 Hestia	& SOFIA G227		    & 131 $\pm$ 22 \ddag	& 0.046	$\pm$ 0.015 \ddag	& 90.5	$\pm$ 3.0 * &	0.47 $\pm$ 0.03* \\
		      & SOFIA G111 \& G227	  & 131 $\pm$ 22 \ddag	  & 0.046	$\pm$ 0.015 \ddag	& 132 $\pm$ 3.7 &	0.96 $\pm$ 0.01 \\
65 Cybele	& Spitzer SH		    & 237 $\pm$ 4.2 \dag	& 0.071	$\pm$ 0.003 \dag	& 289 $\pm$ 15  &	0.97 $\pm$ 0.02  \\
		      & Spitzer SL 		      & 237 $\pm$ 4.2 \dag	  & 0.071	$\pm$ 0.003 \dag	& 292 $\pm$ 15  &	0.94 $\pm$ 0.01  \\
87 Sylvia	& Spitzer SH \& LH	    & 253 $\pm$ 3.0 \ddag	& 0.046	$\pm$ 0.004 \ddag	& 246	$\pm$ 12 &	0.96 $\pm$ 0.01 \\
		      & Spizter SL 		      & 253 $\pm$ 3.0 \ddag	  & 0.046	$\pm$ 0.004 \ddag	& 263 $\pm$ 13 &	0.98 $\pm$ 0.01 \\
		      & SOFIA G111 \& G227	  & 253 $\pm$ 3.0 \ddag	  & 0.046	$\pm$ 0.004 \ddag	& 267 $\pm$ 7.3 &	0.89 $\pm$ 0.01  \\
140 Siwa	& Spitzer SL 		    & 110 $\pm$ 3.0 \dag	& 0.068	$\pm$ 0.004 \dag	& 92.6 $\pm$ 4.7 &	0.87 $\pm$ 0.01  \\
		      & SOFIA G111 \& G227	  & 110 $\pm$ 3.0 \dag	  & 0.068	$\pm$ 0.004 \dag	& 98.4 $\pm$ 2.7  &	0.76 $\pm$ 0.01 \\
190 Ismene	& SOFIA G227		    & 159 			& 0.066				& 180 $\pm$ 5.3  &	0.84 $\pm$ 0.02 \\
		      & SOFIA G111		      & 159			& 0.066				& 174 $\pm$ 5.2 &	0.75 $\pm$ 0.01 \\
267 Tirza 	& Spitzer SL \& LL	    & 56.0 $\pm$ 0.6 \ddag	& 0.051	$\pm$ 0.008 \ddag 	&   55.1 $\pm$ 2.7 & 0.91 $\pm$ 0.01 \\
269 Justitia	& Spitzer SL		& 50.7 $\pm$ 0.2 \ddag	& 0.061	$\pm$ 0.007 \ddag	& 50.5 $\pm$ 2.5 &	0.84 $\pm$ 0.01 \\
336 Lacadiera	& Spitzer SL 		& 63.4 $\pm$ 11 \ddag 	& 0.054	$\pm$ 0.023 \ddag	& 69.9 $\pm$ 3.5 &	0.99 $\pm$ 0.01  \\
368 Haidea 	& Spitzer SL 		    & 69.3 $\pm$ 0.2 \ddag	& 0.038	$\pm$ 0.007 \ddag	& 66.7 $\pm$ 3.3  &	0.88 $\pm$ 0.01 \\
		      & Spitzer LL		      & 69.3 $\pm$ 0.2 \ddag	& 0.038	$\pm$ 0.007 \ddag 	& 67.1 $\pm$ 3.4 &	0.84 $\pm$ 0.01 \\
944 Hidalgo	& Spitzer SL \& LL	    & 38.0			& 0.060				& 45.6 $\pm$ 2.3  & 	0.77 $\pm$ 0.01 \\
1268 Libya	& Spitzer SL		    & 96.7 $\pm$ 1.2 \ddag	& 0.043	$\pm$ 0.008 \ddag 	& 101 $\pm$ 5.1 &	0.91 $\pm$ 0.01  \\
		      & Spitzer LL		      & 96.7 $\pm$ 1.2 \ddag	& 0.043	$\pm$ 0.008 \ddag	& 95.8 $\pm$ 4.8  &	0.91 $\pm$ 0.01 \\
1702 Kalahari	& Spitzer SL		& 34.6 $\pm$ 0.1 \ddag	& 0.065	$\pm$ 0.008 \ddag	& 34.7 $\pm$ 1.7 &	0.85 $\pm$ 0.01 \\
		      & Spitzer LL		      & 34.6 $\pm$ 0.1 \ddag	& 0.065	$\pm$ 0.008 \ddag	& 33.7 $\pm$ 1.7 &	0.75 $\pm$ 0.01 \\
3552 Don Quixote &	Spitzer LL	    & 19.0			& 0.030			& 18.7 $\pm$ 0.9 &	0.76 $\pm$ 0.01  \\
\enddata
\tablecomments{For each asteroid, literature values for diameter and albedo were taken from JPL Horizons. Observations marked with \dag{} are based on IRAS data (\cite{IRAS}) and those marked with \ddag{} are based on NEOWISE (\cite{Mainzer_NEOWISE}) measurements. The diameter and albedo values cited in the JPL Horizons database for 190 Ismene were sourced from personal communication with J. G. Williams and did not have associated error estimates. Similarly, the diameter and albedo values of 944 Hidalgo and 3553 Don Quixote were sourced from \cite{gehrels} and also did not include error estimates.  Note that a lack of contemporaneous photometry for the June G227 observations of 46 Hestia (starred) resulted in a fit with an asteroid diameter that does not agree with literature values, as these observations could not be scaled to account for potential slit losses.}
\end{deluxetable*}

\section{Results}

\subsection{Emissivity Spectra and Thermal Modeling}

The results of the NEATM modeling are summarized in Table \ref{tab:NEATM}. We note good agreement between the reported literature values and NEATM derived diameters, with a few exceptions. First, the June SOFIA G227 observations of 46 Hestia do not agree with literature values for diameter. These observations were taken without contemporaneous photometry, so these spectra could not be scaled to account for potential slit losses. Accordingly, the NEATM derived diameter and beaming parameter is lower for this observation than for the other observation of 46 Hestia. For this observation in particular, the June SOFIA G227 observations were converted to emissivity by dividing by the thermal model, then multiplying the resulting emissivity spectrum by a scaling factor to align the June G227 observations that lacked photometry to the July G227 observations. These two G227 observations were averaged together to produce the final G227 spectrum of 46 Hestia, so the June values of beaming parameter and diameter are likely a reflection of observation circumstances (e.g. slit misalignment) rather than reflective of a true physical property of 46 Hestia.

Another exception is 65 Cybele. Using the same Spitzer dataset, \cite{licandro_2011} found 65 Cybele had an estimated diameter of 290 $\pm$ 5 km. Our diameter estimate for 65 Cybele is much larger than the NEOWISE diameter but is consistent with the diameter derived by \cite{licandro_2011}, suggesting the difference between the derived diameter values is due to observing 65 Cybele at different epochs. Temporal variation, including the temporally varying cross section of a rotating asteroid, or differences in data processing, such as rotational or seasonal averaging of thermal fluxes used to estimate diameter, may account for the differences in reported literature diameter and derived diameter for asteroids like 87 Sylvia and 140 Siwa, whose diameter values differ by many standard deviations. This interpretation is further supported by the \cite{marchis2012sylvia} diameter estimate. That work estimated the diameter of 87 Sylvia in the Spitzer SL data to be 272.4 $\pm$ 13.4 km, in agreement with our 263 $\pm $ 13 km diameter derived from the same dataset. Therefore, variations in Sylvia's derived diameter likely reflect varying observation circumstances between observing epochs. The results of our thermal modeling can be further validated by comparing the derived NEATM parameters from previous spectroscopic observations and independently computed thermophysical models of the same Spitzer archival data. In \cite{mommert2014DQ}, the diameter of 3552 Don Quixote is estimated as 18.4 $\pm$ 0.04 km. The diameter estimates for these asteroids are in agreement with the diameter estimates of the thermal model presented in this paper. \cite{vernazza2013} presents an emissivity spectrum of 368 Haidea, that paper does not does not report an estimated size. Similarly, diameter is not a varied parameter in the thermophysical modeling of 944 Hidalgo in \cite{lowry2022t}. 

\begin{figure}[!ht]

\gridline{\fig{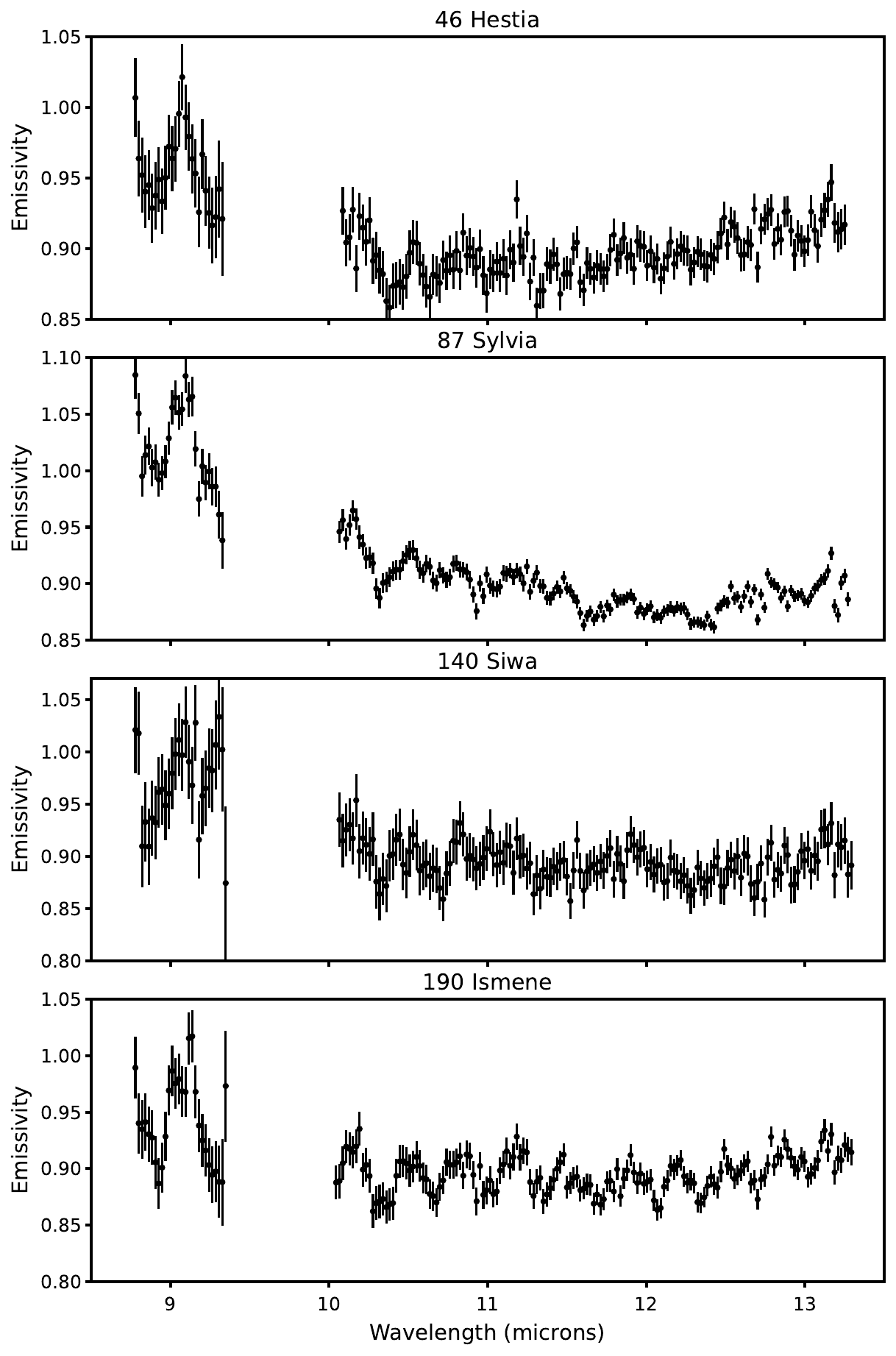}{0.5\textwidth}{(a)}
          \fig{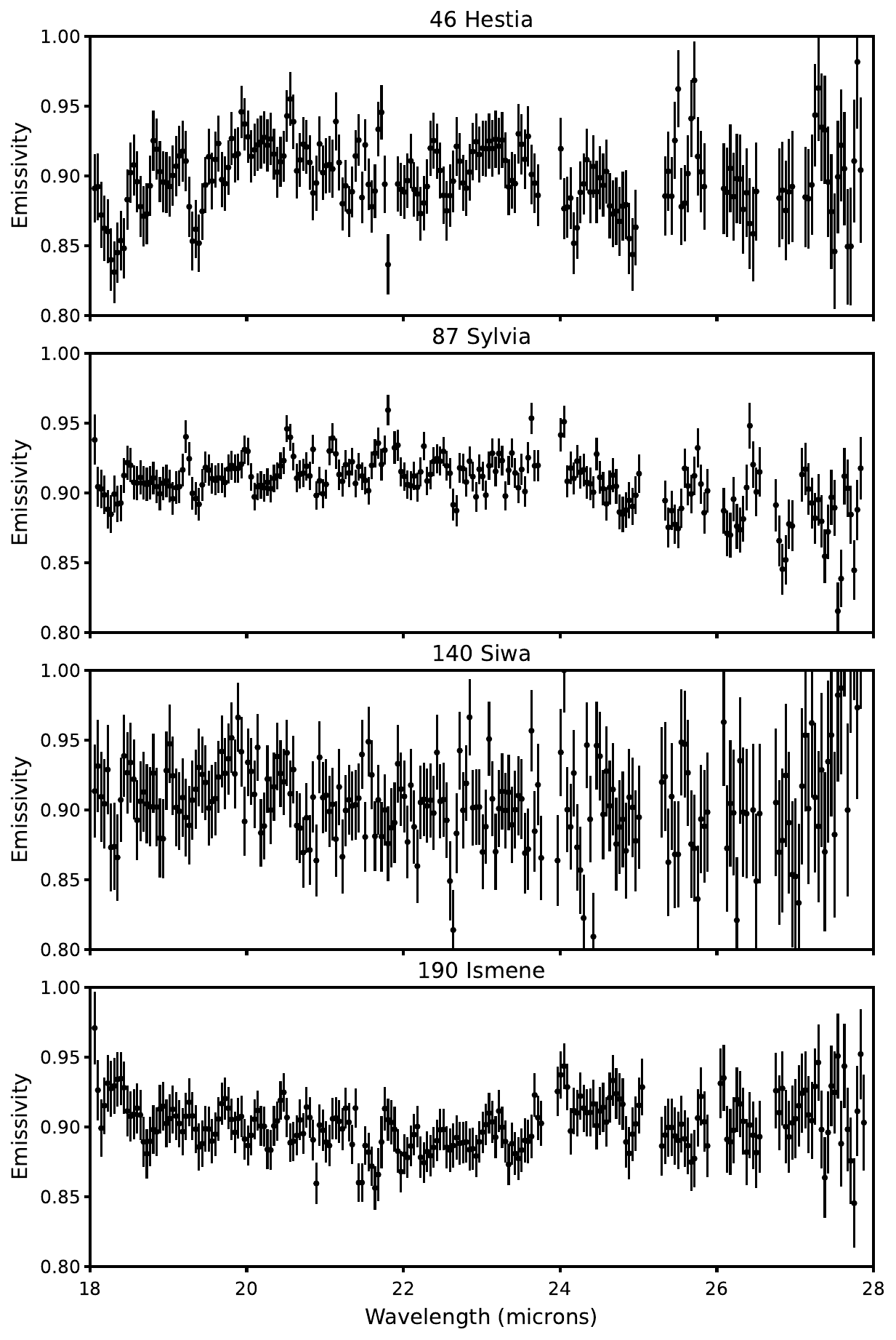}{0.5\textwidth}{(b)}
          }
\caption{Mid-infrared spectra of asteroids (46) Hestia, (87) Sylvia, (140) Siwa, and (190) Ismene as observed by the SOFIA FORCAST instrument, using the G111 (panel a) and G227 (panel b) grisms. Note that we do not display data points for which atmospheric transparency is low ($<0.9$).    \label{fig:SOFIA_full}}
\end{figure}

\begin{figure}[!ht]
\gridline{\fig{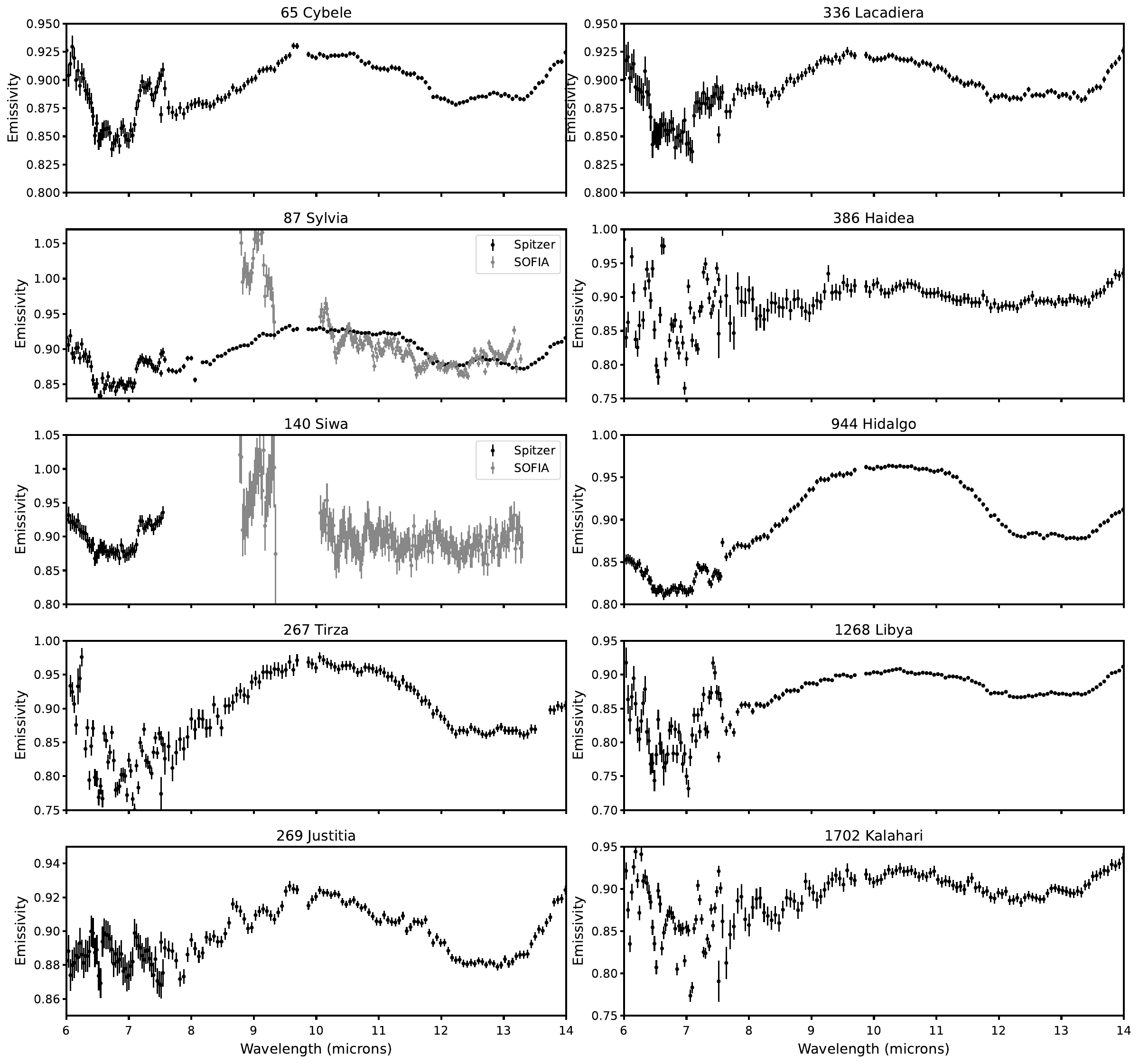}{1\textwidth}{}}
\caption{Mid-infrared spectra of asteroids (65) Cybele, (87) Sylvia, (140) Siwa, (267) Tirza, (269) Justitia, (336) Lacadiera, (386) Haidea, (944) Hidalgo, (1268) Libya, and (1702) Kalahari from 6-14 $\mu$m as observed by the Spitzer IRS instrument in SL mode. For (87) Sylvia and (140) Siwa, we also overplot the SOFIA FORCAST G111 observations scaled to the same average emissivity for comparison. See also \cite{licandro_2011, marchis2012sylvia, vernazza2013}, and \cite{lowry2022t} for their analyses of the same Spitzer IRS spectra of (65) Cybele, (87) Sylvia, (386) Haidea, and (944) Hidalgo respectively. \label{fig:Spitzer_short}}
\end{figure}

\begin{figure}[!ht]
\gridline{\fig{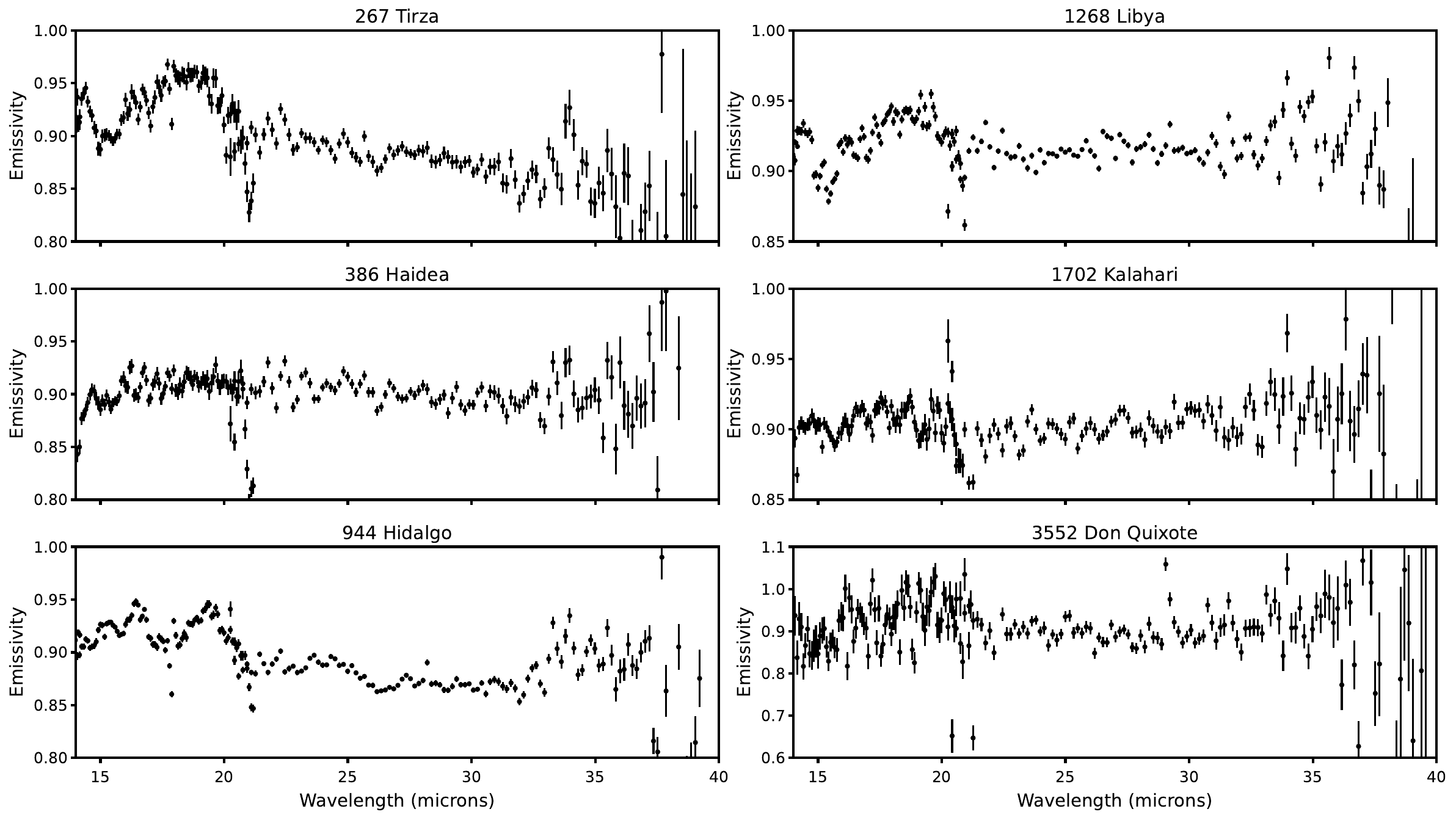}{1\textwidth}{}}
\caption{Mid-infrared spectra of asteroids (267) Tirza, (386) Haidea, (944) Hidalgo, (1268) Libya, (1702) Kalahari, and (3552) Don Quixote from 14-40 $\mu$m as observed by the Spitzer IRS instrument in LL mode. See also \cite{vernazza2013, lowry2022t}, and \cite{mommert2014DQ} for their analyses of the same Spitzer IRS spectra of (368) Haidea, (944) Hidalgo, and (3552) Don Quixote respectively.}  \label{fig:Spitzer_long}
\end{figure}

\subsection{Feature Characterization}

We present the complete SOFIA spectra (8-14 $\mu$m \& 18-28 $\mu$m) in Figure \ref{fig:SOFIA_full}. The results of our reduction of the Spitzer Space Telescope spectra (5 - 40 $\mu$m) are shown in Figures \ref{fig:Spitzer_short} and \ref{fig:Spitzer_long}. In general, the Spitzer spectra tend to show  a sharp emissivity dip at short wavelengths, followed by the broad 10-$\mu$m silicate feature in emission. At longer wavelengths, the spectra tend to show additional silicate features around 15 and 20 $\mu$m followed by a flat or linear continuum. To describe the general shapes of the features we employ the terminology used in \cite{kelley2017mid} to compare the mid-infrared spectra of comets and Jupiter Trojans: the broad 10-$\mu$m features are characterized as trapezoidal, rounded, or present. 

Among our sample, we see trapezoidal 10-$\mu$m features in the Spitzer spectra of 65 Cybele, 87 Sylvia, 267 Tirza, and 944 Hidalgo. These asteroids all show sharp shoulders (e.g. a change in slope from the relatively flat slope of the 10-$\mu$m plateau to a decreasing slope) near 11 $\mu$m, bounding the long wavelength edge of the emission features. 

We see rounded 10-$\mu$m features in the spectra of 269 Justitia, 336 Lacadiera, 1268 Libya, and 1702 Kalahari. These asteroids show 10-$\mu$m features bounded by dips at 8 and 12 $\mu$m like the trapezoidal asteroids, but the shoulder at 11 $\mu$m is less distinct but still present. 386 Haidea also has a rounded 10-$\mu$m feature, but unlike the previous examples, 386 Haidea lacks a peak or shoulder near 11 $\mu$m. Instead, the spectrum of 386 Haidea begins to turn downwards at 10.55 $\mu$m. 

Without complete coverage of the short wavelength end of the 10 $\mu$m feature, it is harder to classify the shapes of the 10-$\mu$m feature in the SOFIA data. The general decrease in emissivity from 10 $\mu$m to 12.5 $\mu$m is evidence of the 10-$\mu$m silicate emission feature in the SOFIA spectrum of 87 Sylvia. Similarly, the higher average emissivities from 10 to 11 $\mu$m and presence of the 12.5 $\mu$m dip in the SOFIA spectra of both 87 Sylvia and 140 Siwa (which is associated with and defines the long wavelength edge of the 10 $\mu$m feature) suggests that the 10 $\mu$m feature is present in the spectra of both these asteroids. For 190 Ismene, interpretation of spectral features is complicated by the presence of a systematic, high frequency 'wiggle' in wavelength space present in the G111 spectrum (8 - 14 $\mu$m), which shows a series regular peaks separated by $\sim 0.34$ $\mu$m across the short-wavelength portion of the spectrum. The other SOFIA spectra may be similarly affected by high frequency noise. For 46 Hestia, we see an increase in emissivity from 10 to 12.5 $\mu$m and absence of a clear high wavelength bound on the 10 $\mu$m feature suggesting that the 10 $\mu$m feature is not present in emission at all. From 10 to 13 $\mu$m, the spectrum of 87 Sylvia is consistent (aside from high frequency noise that may be atmospheric) in both the SOFIA and Spitzer data. We note a discrepancy in the Spitzer and SOFIA spectra at the shortest SOFIA wavelengths ($\sim$ 9 $\mu$m), where the SOFIA spectrum of 87 Sylvia has a much higher emissivity than the Spitzer data. We see similar extremely high emissivities at the shortest wavelengths (5 - 6 $\mu$m) of the Spitzer SL data as well, suggesting that the thermal model struggles to match the data in regions of particularly low flux (e.g. at the shortest wavelengths for any particular instrument). Given the higher signal-to-noise ratio and more complete wavelength coverage, we defer to the Spitzer spectra of 87 Sylvia in our analyses.

To determine the precise locations of spectral features in each asteroid spectrum (See Table \ref{tab:feat_location}), we used a Monte Carlo estimation method. For each asteroid in our sample, we generated 1000 synthetic asteroid spectra by drawing emissivity values from a Gaussian distribution centered around the emissivity value of each data point with a standard deviation equal to the error for that point. Then, for each synthetic spectrum, we fit a fourth-order polynomial spline to smooth the data, resulting in 1000 smoothed splines for each spectrum. Using a polynomial spline allows local extrema the spectrum to be efficiently identified as the derivatives of polynomials are straightforward to compute. To identify a particular local maximum (minimum), we first identify a range over which to determine the maximum (minimum) visually, then calculate the local extrema for each spline by determining where its derivative is 0. In the case where there are multiple points with a derivative of 0, we evaluate the spline at each point and take the wavelength corresponding to the maximum (minimum) value as the location of the local maximum (minimum). Given the locations of the spline maxima (minima) in wavelength space, we then fit a Gaussian to these measurements to get an average feature position (corresponding to the mean) and error (corresponding to the standard deviation) of the distribution of measurements. Repeating this process for different local extrema allows us to measure the position of various features in the spectrum. This procedure is similar to the feature measurement process described in \cite{martin-trojan} and (in further detail, in \cite{martin_diss}).

\begin{figure}[!ht]
\gridline{\fig{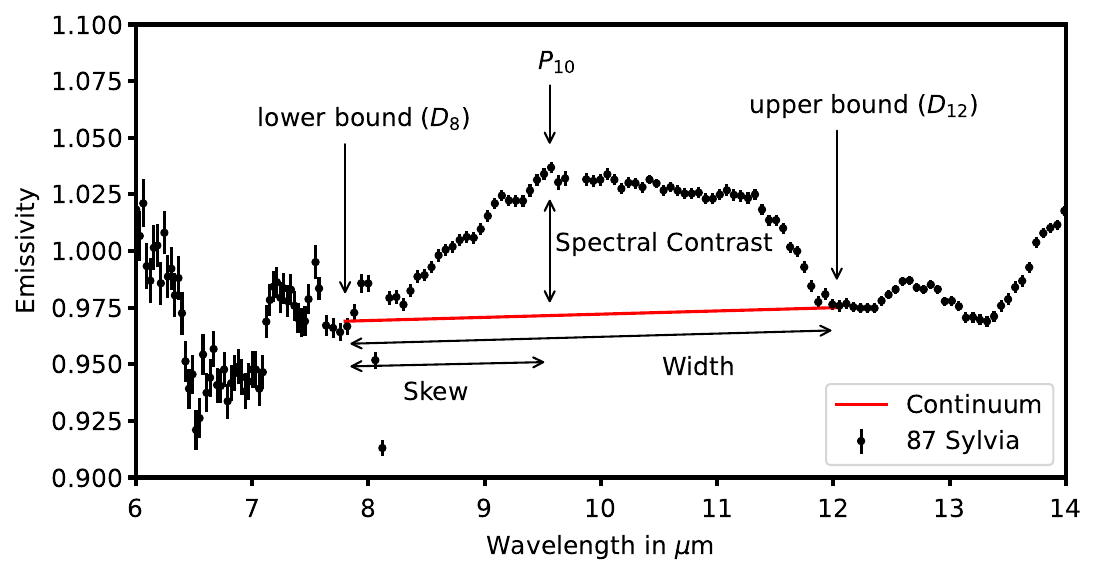}{0.5\textwidth}{}}
\caption{Graphical explanation of the band parameters of the 10-$\mu$m feature, with the spectrum of (87) Sylvia, normalized to an average emissivity of 1, as an example. Band parameters are measured as described in \cite{martin_diss}, except for skew (see text).  \label{fig:band_param}}
\end{figure}

In addition to measuring the locations of various features, we also use the methods described in \cite{martin2022_olivine, martin-trojan} to characterize the shape of the 10-$\mu$m silicate emission feature to enable direct comparison of the spectral features of our Main Belt sample to the Trojans (See Figure \ref{fig:band_param}). In \cite{martin-trojan}, local minima near 8 and 12 $\mu$m are used to define the bounds of the silicate emission feature. In particular, the minimum at 8 $\mu$m is bounded on the lower wavelength side by a sharp dip in emissivity which identifies the edge of the feature. For two asteroids (368 Haidea and 1702 Kalahari) in our sample, we identify the feature edge closer to 7 $\mu$m than 8 $\mu$m because of the location of this sharp dip. Once the bounds are identified using the procedure described previously to locate local extrema, the mean location of the 8- and 12-$\mu$m features are used to define endpoints for a linear continuum, which can be subtracted from the spectrum to isolate the spectrum of the 10-$\mu$m feature itself. We parameterize the feature by its width, continuum slope, spectral contrast, and skew (see Table \ref{tab:10micron}) as follows:
\begin{itemize}
    \item Width is measured by subtracting the locations of the 12- and 8-$\mu$m bounds. 
    \item Continuum slope is given by the slope of the linear continuum passing between the 12- and 8-$\mu$m bounds.
    \item To obtain spectral contrast, we normalize the spectrum to a mean emissivity of 1, then determine the location of the continuum removed peak near 9.8 $\mu$m. Subtracting the emissivity at this peak by the predicted emissivity from the continuum gives a difference in emissivity that can be multiplied by 100 to give a percentage increase in emissivity
    \item We measure skew as a proportion, dividing the distance from the 9.8-$\mu$m peak to the 8-$\mu$m bound by the width of the feature. A left-skewing feature has a skew $<0.5$, indicating that the continuum-removed peak falls at wavelengths shorter than the midpoint of the bounds, whereas a right-skewing feature has a skew $>0.5$, indicating that the continuum-removed peak falls at wavelengths longer than the midpoint of the bounds. For example, the skew of 87 Sylvia is 0.44, indicating that the peak of the 10 $\mu$m feature occurs just shortwards of the midpoint of the continuum line (See \ref{fig:band_param}). Note that this method for measuring skew matches the one described for measuring the skew of laboratory spectra in \cite{martin2022_olivine} but is different from the one that \cite{martin-trojan} uses to measure the skew of asteroid spectra.
\end{itemize} 
All the asteroids we examine are skewed towards shorter wavelengths, that is, they have skews $<0.5$. We also note that for many of the asteroids in our sample, the continuum slopes are flat: this is in contrast with the Trojans in \cite{martin-trojan}, which in general have positive slopes (e.g. increasing emissivity with increasing wavelength). 944 Hidalgo, 1268 Libya, and 1702 Kalahari are exceptions to this general trend: these asteroids all have positive continuum slopes significant at the 1-$\sigma$ level. We do not see any asteroids with statistically significant negative continuum slopes. In terms of spectral contrast, the contrasts we calculate are comparable to those seen in the Trojans (\cite{licandro_2011, martin-trojan}).

\begin{deluxetable*}{lccccccc}[ht]
\tablenum{5}
\tablecaption{Location of Common Spectral Features \label{tab:feat_location}}
\tablewidth{0pt}
\tablehead{
Asteroid & $D_{5.5}$ & $P_6$ &  $D_7$ &	$P_{7.5}$	& $D_8$ &	$P_9 / S_9$&	$P_{10}$	\\
& ($\mu$m) & ($\mu$m)& ($\mu$m)& ($\mu$m)& ($\mu$m)& ($\mu$m)& ($\mu$m)\\
}
\startdata
46 Hestia & -	& - & - & - & - & $9.06\pm0.01$* & - \\
65 Cybele & $5.68\pm0.09$ & $5.97\pm0.01$ & $6.88\pm0.06$ & $7.38\pm0.10$ & $7.90\pm0.10$ & X & $9.73\pm0.05$ \\ 
87 Sylvia & $5.77\pm0.06$ & $6.04\pm0.06$ & $6.62\pm0.16$ & $7.54\pm0.06$ & $7.80\pm0.04$ & $9.15\pm0.05$ & $9.56\pm0.02$ \\
140 Siwa & $5.73\pm0.02$ & $5.95\pm0.01$ & $6.77\pm0.17$ & $7.46\pm0.07$ & X* & $9.14\pm0.24$* & X*\\
190 Ismene & - & - & - & - & - & $9.12\pm0.02$* & - \\
267 Tirza & 5.81$\pm$0.03 & 6.02 $\pm$ 0.01  & 7.08 $\pm$ 0.03 & X & 7.80 $\pm 0.10 $  & 9.35 $\pm$ 0.03 & 9.84 $\pm$ 0.15  \\
269 Justitia & $5.78\pm0.03$ & $6.72\pm0.08$ & $7.00\pm0.05$ & X & $7.85\pm0.02$ & X & $9.66\pm0.13$ \\
336 Lacadiera & $5.65\pm0.12$ & $5.95\pm0.07$ & $6.60\pm0.12$ & $7.51\pm0.09$ & $7.63\pm0.08$ & $9.14\pm0.10$ & $9.77\pm0.06$ \\
368 Haidea &  X & X & $6.69 \pm 0.01$ & $7.59 \pm 0.08$ & $8.18 \pm 0.05$ & $9.26 \pm 0.02$ & $10.55 \pm 0.08$ \\
944 Hidalgo & $5.66\pm0.04$ & $5.93\pm0.04$ & $6.52\pm0.05$ & $7.20\pm0.04$ & $7.41\pm0.06$ & $9.20\pm0.04$ & $9.85\pm0.19$ \\
1268 Libya & X & X & $6.82\pm0.19$ & $7.43\pm0.01$ & $7.74\pm0.03$ & X & $10.3\pm0.14$ \\
1702 Kalahari & X & X & $7.07\pm0.01$ & $7.52\pm0.05$ & $7.99 \pm 0.04$ & $9.66 \pm 0.08$ & $10.3 \pm 0.17$ \\
3552 Don Quixote & - & - & - & - & - & - & - \\
\hline
Feature Description & Sharp dip & Plateau & Dip & Plateau & Low bound & Shoulder & Peak of \\
& & & & &  of 10$\mu$m feature & minor peak & 10$\mu$m feature \\
\hline
\hline
Asteroid & $P_{11} / S_{11}$	& $D_{12.5}$ & $	P_{13}$ & $	P_{14.5}$ &$	P_{17}$& $	P_{20}$ & Shape of\\
& \\
 & ($\mu$m)& ($\mu$m)& ($\mu$m)& ($\mu$m)& ($\mu$m)& ($\mu$m) & 10 $\mu$m feature\\
\hline
46 Hestia & X* & X* & X* & - & - & $20.74\pm0.57$* & X\\
65 Cybele & $11.68\pm0.12$ & $12.27\pm0.07$ & $13.30\pm0.04$ & $14.23\pm0.03$ & $17.66\pm0.01$ & - & Trapezoidal \\
87 Sylvia & $11.2\pm0.07$ & $12.03\pm0.11$ & $12.63\pm0.03$ & $15.00\pm0.31$ & $17.39\pm0.35$ & $20.13\pm0.45$ & Trapezoidal \\
140 Siwa  & $10.9\pm0.09$* & $12.45\pm0.11$* & $11.91\pm0.15$* & - & - & $20.07\pm0.46$* & Present \\
190 Ismene & X?* & X?* & X?* & - & - & X* & X? \\
267 Tirza & 11.23 $\pm$ 0.07  & 12.67 $\pm$ 0.10 &12.97 $\pm$ 0.03 &14.33 $\pm$ 0.04  & 16.7 $\pm$ 0.06 & 18.97 $\pm$ 0.21 &   Trapezoidal \\
269 Justitia & $11.72\pm0.07$ & $12.82\pm0.07$ & X & - & - & - & Rounded \\
336 Lacadiera & $11.5\pm0.03$ & $11.97\pm0.11$ & $12.51\pm0.12$ & - & - & - & Rounded \\
368 Haidea & X & $12.18\pm0.04$ & $12.64 \pm 0.21$ &  $14.76 \pm 0.08$ & $16.93 \pm 0.718$ & X & Rounded \\
944 Hidalgo & $11.12\pm0.04$ & $12.28\pm0.06$ & $12.88\pm0.04$ & X & $16.47\pm0.07$ & $19.38\pm0.06$ & Trapezoidal \\
1268 Libya & $11.49\pm0.05$ & $12.30\pm0.04$ & $13.00\pm0.08$ & $14.27\pm0.01$ & X & $19.14\pm0.06$ & Rounded \\
1702 Kalahari &  $11.5 \pm 0.16$ & $12.27 \pm 0.05$ & $12.93\pm0.15$ & $15.11 \pm 0.30$ & $17.93\pm0.85$ & X& Rounded\\
3552 Don Quixote & - & - & - & - & $16.44\pm0.53$ & $20.65\pm0.01$ & - \\
\hline 
Feature Description & Shoulder or  & High bound & Minor & Peak  & Broad & Broad &  \\
& minor peak & of 10-$\mu$m feature & peak & & peak  & peak   & \\
\enddata
\tablecomments{Positions and locations of various features common to the mid-IR spectra of the asteroids analyzed. Measurements based on Spitzer Space Telescope except those marked with a star (*) indicating observations from the SOFIA spectra. The letters P, D, and S indicate whether the feature is a peak, dip, or shoulder, while the corresponding number gives the \textit{approximate} position of the feature. Each feature is also briefly described. An X indicates the feature was not present in the spectrum while a dash (-) indicates there was no coverage of the relevant wavelength region. The X? symbol, used only for (190) Ismene indicates that while this wavelength region was covered, systematic issues in the spectrum complicates the interpretation of individual features.}
\end{deluxetable*}

\begin{deluxetable*}{lccccc}[ht]
\tablenum{6}
\tablecaption{Characterization of the 10-$\mu$m feature \label{tab:10micron}}
\tablewidth{0pt}
\tablehead{
	& Width	& Continuum Slope & 	Continuum Removed Peak &		& Spectral Contrast \\
Asteroid	& ($\mu$m) &	(Emissivity/$\mu$m)&	($\mu$m)&	Skew &	(\%) \\
}
\startdata
65 Cybele	& 4.37 $\pm$ 0.12	& 0.001 $\pm$ 0.002	& 9.68 $\pm$ 0.04	& 0.41 $\pm$ 0.07   & 5.8 $\pm$ 0.6 \\
87 Sylvia	& 4.23 $\pm$ 0.12	& 0.003 $\pm$ 0.001	& 9.68 $\pm$ 0.14	& 0.44 $\pm$ 0.10	& 6.1 $\pm$ 0.6 \\
267 Tirza	& 4.87 $\pm$ 0.14 & 0.005 $\pm$ 0.005 & 9.81 $\pm$ 0.15 &0.41 $\pm$	0.09 & 12 $\pm$ 2.0 \\
269 Justitia &	4.87 $\pm$ 0.07	& 0.002 $\pm$ 0.001 & 9.64 $\pm$ 0.15 & 0.36 $\pm$ 0.09 & 5.1 $\pm$ 0.8 \\
336 Lacadiera &	4.34 $\pm$ 0.14	& 0.001 $\pm$ 0.003	& 9.63 $\pm$ 0.17	& 0.46 $\pm$ 0.11	& 4.6 $\pm$ 0.9 \\
368 Haidea	& 3.9 $\pm$ 0.06 & 0.004 $\pm$ 0.005 & 9.26 $\pm$ 0.02 & 0.25 $\pm$ 0.06 & 6.5 $\pm$ 2.1 \\
944 Hidalgo	& 4.87 $\pm$ 0.08	& 0.009 $\pm$ 0.002	& 9.81 $\pm$ 0.17	& 0.49 $\pm$ 0.08	& 11 $\pm$ 0.7 \\
1268 Libya	  & 4.56 $\pm$ 0.05	& 0.010 $\pm$ 0.002	& 9.61 $\pm$ 0.26	& 0.41 $\pm$ 0.14	& 6.7 $\pm$ 0.7 \\
1702 Kalahari &  $4.63\pm0.05$ & $0.014 \pm 0.004$ & $9.43\pm0.12$ & $0.39 \pm 0.07$ & 7.3$\pm$ 1.8  \\
\enddata
\tablecomments{Band parameters of the 10 $\mu$m feature. Band parameters are measured as described in \cite{martin-trojan}, except for skew (see text).}
\end{deluxetable*}

\section{Discussion}

To identify the mineralogies present on the surfaces of these asteroids, we follow the approach of \cite{martin-trojan}. That work used the characteristic wavelengths of individual peaks and dips superimposed on the broad 10-$\mu$m silicate emission feature to argue that the spectra of a collection of Trojans showed evidence of the presence of forsteritic olivine and enstatite (pyroxene). \cite{martin2022_olivine} and \cite{MARTIN_pyroxene} also demonstrated that while the spectral contrast of diagnostic silicate features varies with regolith porosity, the locations of many features in wavelength space does not vary. Therefore, these features can be used to identify silicate mineralogy independently of regolith porosity, with the appearance of multiple peaks associated with the same mineral interpreted as strong evidence of the presence of that mineral in the asteroid's spectrum. In addition to using the laboratory studies of \cite{martin2022_olivine, MARTIN_pyroxene}, we also employ previous work including laboratory measurements of the absorption coefficients of olivines and pyroxenes (e.g. \cite{koike1993_olvpyr, koike2003olivine, chihara2002pyroxene}). To compare the absorption coefficients measured in those papers and the emission features considered in this work, we make the following assumptions. First, we assume that the contribution of surface scattering to the spectral appearance of the asteroids in our sample is negligible, as evidenced by the appearance of the 10-$\mu$m feature in emission and the lack of a Christiansen feature (CF), both of which are associated with the transition from the primarily surface scattering regime to the volume scattering regime (\cite{martin2022_olivine, MARTIN_pyroxene}. We also assume that we can relate emission and absorption features according to Kirchhoff's law, an assumption commonly used among laboratory studies of mid-IR spectroscopy (e.g. \cite{SALISBURY1991}) that predicts that the positions of spectral features are unchanged when considered in terms of emissivity rather than reflectance. Like \cite{martin-trojan}, we assume the emission features we observe are linearly additive--that is, that the characteristic features of olivine and pyroxene mixtures can be identified by their diagnostic peaks superimposed on the broader 10- and 20-$\mu$m peaks. 

\begin{figure}
    \centering
     \gridline{\fig{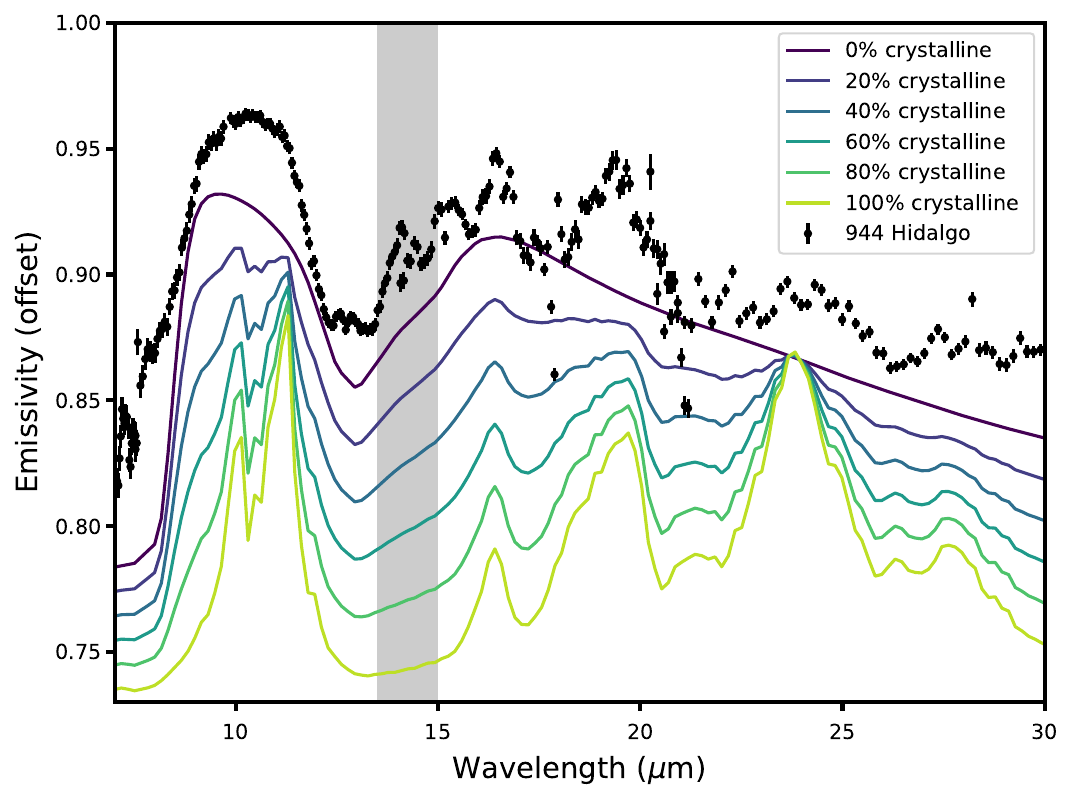}{0.8\textwidth}{}}

    \caption{Modeled mixtures of magnesium rich crystalline and amorphous olivine compared to the spectrum of 944 Hidalgo. Increasing the proportion of amorphous olivine in these mixtures dampens the spectral contrast of the smaller peaks superimposed on the 10-$\mu$m silicate plateau, suggesting a high proportion of amorphous silicates among the asteroids observed with Spitzer and SOFIA. The shaded grey box highlights the location of a known artefact in the SL Spitzer data (the SL teardrop). Spectra of crystalline silicates are sourced from \cite{koike2003olivine} and \cite{chihara2002pyroxene}, spectra of amorphous silicates are derived from optical constants from \cite{jager1994steps, dorschner1995steps} and \cite{jager2003steps} (see text for modeling details and silicate compositions).}
    \label{fig:amorph-olv}
\end{figure}

\begin{figure}
    \centering
     \gridline{\fig{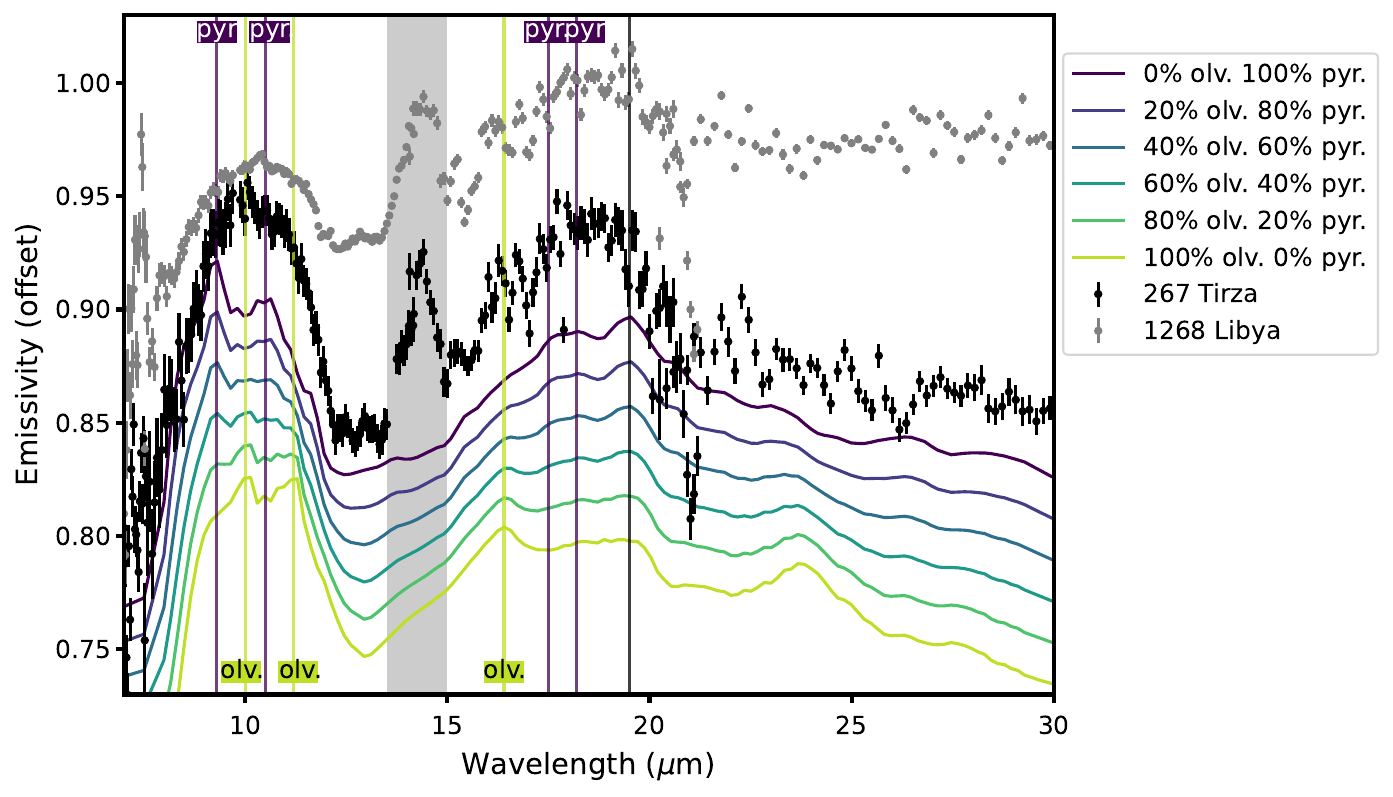}{\textwidth}{}}

    \caption{Modeled mixtures of magnesium rich pyroxenes and olivines plotted alongside the spectrum of 267 Tirza and 1268 Libya. The addition of pyroxene to the olivine mixtures changes the behavior of the spectrum at long wavelengths, broadening this emission feature from the distinct emission peak(s) seen in the spectra of olivine and pyroxene alone. Spectral features of olivine are marked in light green and of pyroxenes are marked in dark purple to highlight commonalities with the asteroid spectra. The feature at 19.5 $\mu$m is common to both olivine and pyroxene. The spectrum of 267 Tirza in particular shows evidence of both olivine and pyroxene. All mixtures shown here contain both amorphous and crystalline silicates mixed in a 3:1 ratio, while the proportion of olivine and pyroxene is varied. The shaded grey box highlights the location of a known artefact in the SL Spitzer data (the SL teardrop). Spectra of crystalline silicates are sourced from \cite{koike2003olivine} and \cite{chihara2002pyroxene}, spectra of amorphous silicates are derived from optical constants from \cite{jager1994steps, dorschner1995steps} and \cite{jager2003steps} (see text for modeling details and silicate compositions).}
    \label{fig:olv-pyr}
\end{figure}

We identify crystalline olivine, amorphous olivine, crystalline pyroxene, and amorphous pyroxene as likely majority components contributing to the overall appearance of the 10- and 20-$\mu$m silicate emission features. Both olivine and pyroxene are names given to classes of silicate minerals that can exhibit varying bulk proportions of Mg and Fe in solid solution. For olivine, forsterite refers to the the Mg-rich endmember and for fayalite refers to the Fe-rich endmember. For pyroxene, the Mg-rich endmember is enstatite and the Fe-rich endmember is ferosilite. In general, unless otherwise noted, we use the spectra of the magnesium-rich endmembers for each silicate. Our choice to use the magnesium-rich endmembers was influenced by the observation that increasing Mg\# in both olivine and pyroxene tends to shift individual features to shorter wavelengths (e.g. \cite{koike2003olivine, chihara2002pyroxene}). As the spectral features of the asteroids we examined in this paper tend to skew towards shorter wavelengths, we selected the most magnesium-rich olivine or pyroxene available with sufficient resolution in the 10-$\mu$m region to be our example endmember. \cite{martin-trojan} also explores how the addition of silicates with lower Mg\# acts to change the skew of the 10-$\mu$m feature as applied to the spectrum of 617 Patroclus-Menoetius. We produce simple linear mixture models using these four silicates to explore how the inclusion of these components affect the spectral shapes of the 10- and 20-$\mu$m silicate emission features and provide a visual comparison between these silicate mixtures and the asteroid spectra considered in this work (Figures \ref{fig:amorph-olv} and \ref{fig:olv-pyr}). To generate these mixing models, we model the predicted emission for amorphous silicates starting from the optical constants for a silicate grain with a radius of 5 $\mu$m following the methods of \cite{lisse2006spitzer}. We used optical constants reported in \cite{jager1994steps, dorschner1995steps} for amorphous olivine (Mg\# = 1.0) and \cite{jager2003steps} for amorphous pyroxene (Mg\# = 0.95). For crystalline silicates, we use data from \cite{koike2003olivine} and \cite{chihara2002pyroxene} for forsteritic olivine (Mg\# = 1.0) and ortho-enstatite pyroxene (Mg\# = 1.0), respectively. The iron-rich endmembers considered in the discussion of 368 Haidea are 1:1 mixtures of amorphous and crystalline silicates with Mg\# = 0.4 for the amorphous silicates (the lowest Mg \# measured in \cite{jager1994steps, dorschner1995steps, jager2003steps}) and Mg\# = 0.0 for the crystalline silicates (\cite{koike2003olivine, chihara2002pyroxene}).

We then produced linear mixtures of the emission spectra of the spectral endmembers to create a synthetic spectrum. The resulting synthetic emission spectrum was then scaled as in \cite{martin-trojan} to match the spectral contrast of the mixture model to the spectral contrast of the 10-$\mu$m emission feature in the asteroid spectra. These simplified models do not account for all the complexities inherent to asteroidal surfaces including grain size distributions and the addition of non-silicate components, but are used here to illustrate a few key principles contributing to our analysis of the silicate mineralogy of the asteroids.

Following these principles, our results suggest mineralogical similarities between the Trojans and D- \& P-types in the Main Belt. In particular, the peak or shoulder between 11.2 and 11.4 $\mu$m is associated with crystalline olivine, with the range of feature positions associated with the different Mg\# within the crystal structure of olivine (\cite{koike1993_olvpyr, koike2003olivine}). Like \cite{martin-trojan}, we take the presence of the 11-$\mu$m features as evidence for the presence of olivine. In the long wavelength region olivines show additional structure including a broad peak from 16-17 $\mu$m, a broad peak 19-20 $\mu$m and a peak near 24 $\mu$m. The spectrum of 944 Hidalgo shows strong evidence of the presence of olivine with the presence of these peaks in the 20-$\mu$m regions (see Figure \ref{fig:amorph-olv}).  

For some asteroids in our sample, there is evidence of pyroxenes contributing to the shapes of the 10- and 20-$\mu$m silicate emission features. While the spectrum of 944 Hidalgo shows the olivine silicate peaks near 20 $\mu$m most clearly, these peaks less clear for 267 Tirza and 1268 Libya, which instead have a broad emission feature peaking near 19 $\mu$m. This broad peak may be due to the presence of pyroxene either alone or in combination with olivine. Some pyroxenes, particularly amorphous pyroxenes, show a prominent and broad long wavelength peak between 19 and 20 $\mu$m (\cite{chihara2002pyroxene, brucato1999mid}) consistent with the interpretation that there is evidence of pyroxene in the spectra of 267 Tirza and 1268 Libya. At short wavelengths, 267 Tirza shows additional evidence of pyroxene via a 9.3-$\mu$m peak associated with pyroxene (\cite{chihara2002pyroxene}) along with the 11.2 $\mu$m peak that we attribute to olivine.  1268 Libya also shows evidence of pyroxenes at short wavelengths with a prominent peak at 10.5 $\mu$m and a smaller peak near 11.5 $\mu$m (\cite{chihara2002pyroxene}), though the latter can also be interpreted as evidence for fayalitic olivine, which has a peak near 11.4 $\mu$m (\cite{koike2003olivine}). Thus, a mixture of olivines and pyroxenes may be able to simultaneously account for the features in the 10- and 20-$\mu$m regions of the spectra of 267 Tirza and 1268 Libya.

In general, emissivity spectra of crystalline silicates are sharper and more distinct than their amorphous counterparts (e.g. \cite{brucato1999mid, HALLENBECK1998198}). Spectral mixing models predict that increasing the proportion of amorphous olivines and pyroxenes in a mixture acts to dampen the distinctiveness of individual silicate emissivity peaks and act to round the overall shape of the 10 $\mu$m plateau (see Figure \ref{fig:amorph-olv} of this paper and Figure 6 of \cite{martin-trojan}). While we can identify individual silicate peaks and shoulders superimposed on the broad 10-$\mu$m plateau, the spectral contrasts of these features are much smaller than would be expected of a purely crystalline composition, evidence that a high proportion of the silicates in on the surfaces of these asteroids are amorphous, particularly among the asteroids classified as having ``rounded'' 10-$\mu$m features (e.g. 336 Lacadiera, 1268 Libya, and 1702 Kalahari). We also see evidence of amorphous silicates on the surfaces of many asteroids in our sample, in particular, with the presence of a very broad peak near 9.8-$\mu$m (particularly in continuum-removed space) of amorphous silicates that is also found in Trojan spectra (\cite{emery_2006, martin-trojan}).

In general, we see many commonalities in the mid-infrared spectra of the D- \& P-type asteroids in our sample and the Trojans. We see evidence of the 10-$\mu$m silicate emission features in the spectra of all the asteroids we examined, with the exception of 46 Hestia. While 46 Hestia lacks evidence of a 10-$\mu$m feature, we note emission features in the 20-$\mu$m region that may be due to silicates. In both the Trojans (\cite{martin-trojan}) and our sample, we see these features exhibiting both rounded and trapezoidal shapes. The range of spectral contrasts of these features are also comparable to the range of contrasts exhibited by the Trojans: 624 Hektor has an estimated spectral contrast of $\sim 15 \%$ (\cite{licandro_2011}) and 617 Patroclus has a spectral contrast of $\sim 4\%$ (\cite{martin-trojan}), while our sample spans contrasts from $4 - 17 \%$. Since spectral contrast of the 10-$\mu$m feature is associated with high regolith porosity (\cite{martin2022_olivine, MARTIN_pyroxene}), this finding implies similar regolith surface properties on both Trojans and Main Belt asteroids. Previous authors have suggested the suspension of regolith particles within a 'fairy-castle' structure (\cite{emery_2006, mueller2010}) or transparent (in the mid-IR) matrix (\cite{Yang_2013, SULTANA2023115492}) to explain the emission features of Jupiter Trojans. Our results suggest these mechanisms are also relevant to the surfaces of D- \& P-type asteroids outside of the Trojan population. Other features our spectra have in common with the Trojans examined in \cite{martin_diss} include the small emissivity peak at 13 $\mu$m (ex. 87 Sylvia, 65 Cybele, and 944 Hidalgo) in the overall emissivity dip between 12 and 14 $\mu$m, though these features were not attributed to any particular silicate composition. 

However, there are some slight differences between the Main Belt sample presented here and the spectra of the Trojans in \cite{martin-trojan} in terms of the shape of the 10-$\mu$m feature: our Main Belt asteroids have flatter continuum slopes and skew to shorter wavelengths, though we note that the measured continuum slope is dependent on the results of the thermal modeling process. The leftward (e.g. towards shorter wavelengths) skews of the 10-$\mu$m plateau among the asteroids in our sample also contrasts with the Trojans, which skew rightwards (e.g. towards longer wavelengths). In \cite{martin-trojan}, the rightward-skewing 10-$\mu$m feature of the Trojans is noted as a major difference between both spectral mixture models of crystalline and amorphous olivines and pyroxenes as well as simulated emission spectra derived from meteoritic analogs in the ASTER spectral library (\cite{martin_diss, aster}). In that work, \cite{martin_diss} suggests that the rightward skews of Trojan 10-$\mu$m emission plateaus could be caused by additional spectral component responsible for muting the short wavelength silicate peaks. In contrast, the leftward skews of the 10-$\mu$m features of asteroids in our Main Belt sample indicate that the currently unknown physical mechanism responsible for skewing the 10-$\mu$m feature of Trojans to longer wavelengths is not relevant to the asteroids in this sample.

In \cite{martin-trojan} VNIR spectral slope was found to be correlated with the spectral contrast of the 10-$\mu$m feature among the Jupiter Trojans.  Therefore, we might expect the objects in our sample with redder spectral slopes, e.g. the D-types (267 Tirza, 336 Lacadiera, 368 Haidea, 944 Hidalgo, and 1702 Kalahari) and 269 Justitia would all have high spectral contrasts. While the asteroids with the three highest spectral contrasts (944 Hidalgo, 267 Tirza, and 1702 Kalahari) are all D-types (\cite{PDS_tax}), the more moderate spectral contrasts of 336 Lacadiera, 368 Haidea, and 269 Justitia do not align with this expectation. In particular, 269 Justitia's exceptionally red NIR (0.7 - 2.5 $\mu$m) slope of 9\%/100 nm (\cite{hasegawa2021}) exceeds the 5\%/100nm NIR slopes of the average R group Trojan (\cite{emery2010near}). Therefore, if Justitia followed the trend observed in \cite{martin-trojan}, we would expect 269 Justitia to have a prominent 10-$\mu$m feature, but its spectral contrast is comparable to the more moderately sloped P-types. The authors in \cite{hasegawa2021} hypothesize that 269 Justitia's extreme red spectral slope in the VNIR suggests this asteroid originated from the 20 - 30 AU region. Similarly, \cite{martin-trojan} suggests a cometary or Kuiper Belt origin for the Trojans based on their spectral similarity to comets. 269 Justitia's 10-$\mu$m emission feature resembles the rounded 10-$\mu$m emission features of some Trojans, particularly the more moderately sloped LR Trojans, (e.g. 617 Patroclus-Menoetius in \cite{martin-trojan, mueller2010}) supporting a common distant outer Solar System origin for both 269 Justitia and the Trojans. However, the spectrum of 269 Justitia shows additional spectral structure (a series of peaks and dips from 7.5 - 10 $\mu$m superimposed on the short wavelength 10-$\mu$m shoulder) not seen among the Trojans, potentially indicating an additional compositional component present on 269 Justitia that is absent from the Trojans. These features, as well as the low spectral contrast of 269 Justitia's 10-$\mu$m feature complicates the interpretation that 269 Justitia and the Trojans are sourced from the same population. 

We note that two of the D-types in our sample, 944 Hidalgo and 3552 Don Quixote, are unique among our sample population for their high eccentricities ($e>0.5$) and semimajor axes ($a>4.0$) AU. Both 944 Hidalgo and 3552 Don Quixote are extinct comet candidates, with 3552 Don Quixote likely originating from the Jupiter Family Comet (JFC) population (\cite{weissman2002evolution}). The spectra of 944 Hidalgo and 3552 Don Quixote also resemble those of 10P/Tempel 2 and 49P/Arend-Rigaux, two JFCs, with their 10- and 20-$\mu$m silicate emission features \cite{kelley2017mid}. The resemblance of the mid-infrared spectrum of 3552 Don Quixote to comets has also been noted in \cite{mommert2014DQ}. That paper also noted evidence of cometary activity on 3552 Don Quixote including a coma and tail. The resemblance of 944 Hidalgo and 3552 Don Quixote to comet populations in the mid-infrared lends support to the hypothesis that these objects originated in cometary populations.

The spectrum of 368 Haidea represents an exception to the trends described above. The spectrum of 368 Haidea lacks strong evidence of the presence of olivine due to the absence of a peak or shoulder at 11 $\mu$m. 368 Haidea's continuum-removed peak is also shifted to shorter wavelengths resulting in a notably low value for skew (see Table \ref{tab:10micron}). \cite{martin2022_olivine} showed that the appearance of a strong 11 $\mu$m feature is characteristic of olivines and appears in laboratory spectra of olivines regardless of regolith porosity. Therefore, while 368 Haidea has a relatively modest spectral contrast compared to other asteroids in our sample, we should still expect to see an 11 $\mu$m feature if the asteroid had an olivine-rich composition, as the appearance of this feature is independent of porosity. While 368 Haidea's 10 $\mu$m plateau feature has a relatively low spectral contrast, its spectral shape in the 10 $\mu$m region is distinct from other asteroids in our sample with similar spectral contrasts that do show evidence of features at 11 $\mu$m (e.g. 87 Sylvia and 1268 Libya). Therefore we disfavor an explanation that the lack of an 11 $\mu$m feature is due to a difference in regolith porosity. 

\begin{figure}
    \centering
     \gridline{\fig{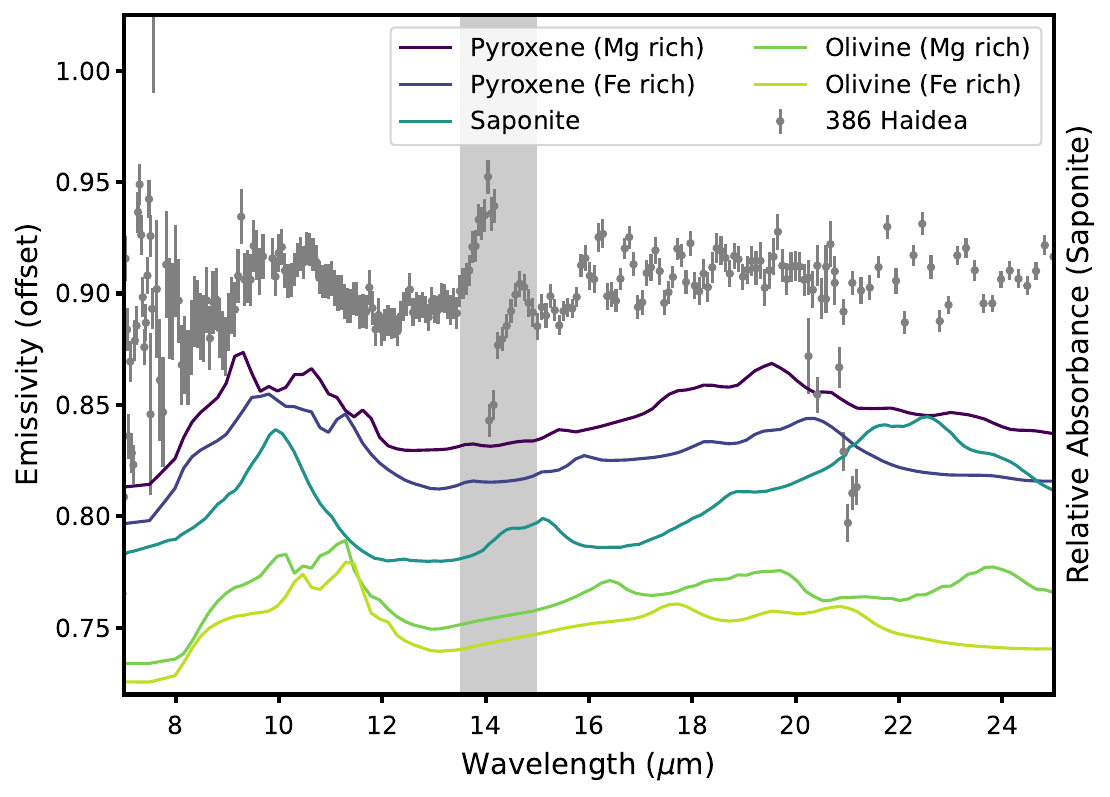}{0.8\textwidth}{}}

    \caption{386 Haidea is distinguished from other asteroid in our sample by its shortward skew and lack of a 11 $\mu$m olivine shoulder, which could be due to a relatively olivine-poor composition. The 10-$\mu$m plateau in pyroxene and phyllosilicate spectra are also skewed to shorter wavelengths and provide a closer match to the shape of 368 Haidea's 10-$\mu$m feature than either forsteritic or fayalitic olivine does. Either of these minerals may be able to account for 386 Haidea's unusual spectrum and could point to a geochemical history different from the olivine-dominated spectra of the other asteroids in our sample. Pyroxene and olivine spectra are modeled as 1:1 mixtures of amorphous and crystalline silicates based on optical constants from \cite{chihara2002pyroxene, jager1994steps, dorschner1995steps} and \cite{jager2003steps}. The saponite spectrum is measured in relative absorbance as derived by \cite{Beck2014} via measurements originally made by \cite{SALISBURY1991}. The shaded grey box highlights the location of a known artefact in the SL Spitzer data (the SL teardrop). Spectra of crystalline silicates are sourced from \cite{koike2003olivine} and \cite{chihara2002pyroxene}, spectra of amorphous silicates are derived from optical constants from \cite{jager1994steps, dorschner1995steps} and \cite{jager2003steps} (see text for modeling details and silicate compositions).}
    \label{fig:skew-haidea}
\end{figure}

\begin{figure}[!ht]
 \gridline{\fig{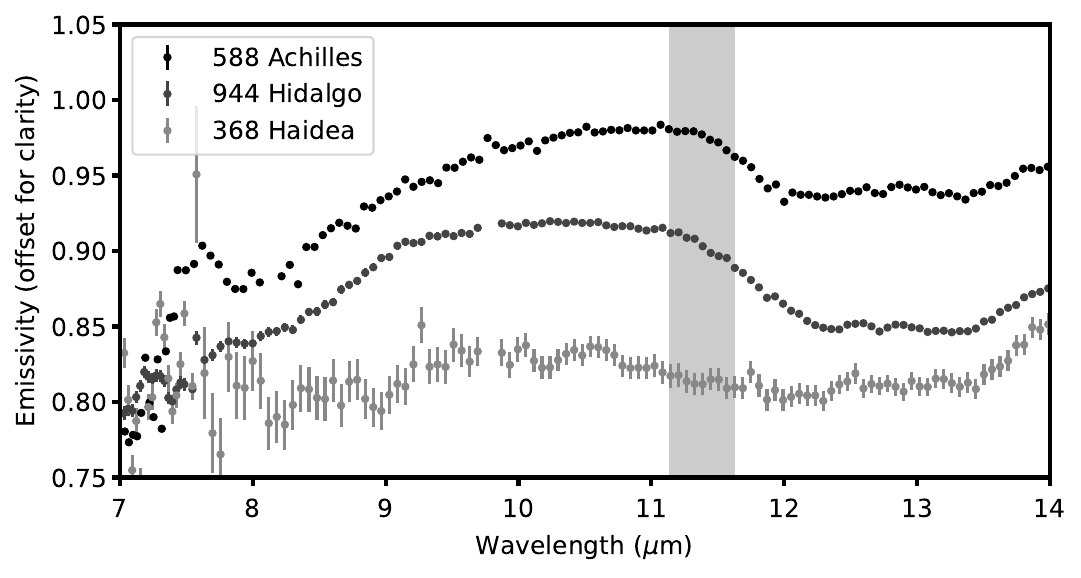}{0.5\textwidth}{}}
 \caption{Comparison of the spectra of 588 Achilles from \cite{martin-trojan}, 944 Hidalgo and 368 Haidea. In the VNIR, 588 Achilles, 944 Hidalgo, and 368 Haidea are all classified as D-types and have similar albedos yet 368 Haidea lacks a strong peak or shoulder in the 11 $\mu$m region indicative of olivine (see highlighted region). Comparing the spectra of 944 Hidalgo and 588 Achilles, we can also see how the asteroids examined in this work have shallower continuum slopes and more leftward skewed 10 $\mu$m feature than the Trojans in \cite{martin-trojan}. \label{fig:compare}}
 \end{figure}

The difference in spectral shape may be due to a difference in composition. 368 Haidea's strongest peaks are located at 9.2 $\mu$m and 10.5 $\mu$m, both prominent peaks associated with pyroxenes, particularly enstatite (\cite{chihara2002pyroxene}), which could be interpreted as evidence for a pyroxene rich composition. Increasing pyroxene content can also explain the shortward skew of 368 Haidea's 10 $\mu$m plateau (The effect of introducing pyroxenes is also demonstrated in Figure \ref{fig:olv-pyr}). However, shortward shifts of the 10-$\mu$m feature, in concert with weakening 11 $\mu$m olivine features have also been associated with increasing degrees of aqueous alteration in meteorites, leading to the alternative conclusion that 368 Haidea may be hydrated (\cite{Beck2014}). As that paper shows, hydrated silicates (in meteorites) can still have quite prominent 10-$\mu$m features, with hydrated minerals often showing a feature near 10 $\mu$m. The hydrated silicate interpretation is supported by the spectral similarity (in the VNIR and mid-IR) of 368 Haidea to the Tagish Lake meteorite noted in \cite{vernazza2013}. The Tagish Lake meteorite has been shown to have widespread evidence of aqueously altered minerals, including phyllosilicates (\cite{zolensky2002,gilmour2019}). Spectra of pyroxenes and the phyllosilicate saponite more closely resemble the spectrum of 368 Haidea than the spectrum of olivine (See Figure \ref{fig:skew-haidea}). Both iron-rich and magnesium-rich olivines do not provide a close spectral match to 368 Haidea, the addition of iron rich olivine (fayalite) in particular skews the 10-$\mu$m feature to even longer wavelengths than forsterite and provides a poorer match. While the pyroxene-rich and phyllosilicate-rich interpretations of 368 Haidea imply different geochemical histories for this asteroid, the lack of evidence for olivine features in the mid-infrared spectrum of 368 Haidea clearly distinguishes it from the Trojans (see Figure \ref{fig:compare}).

Overall, we see that the majority of the primitive asteroids in our sample, with the exception of 368 Haidea, resemble both each other and the Trojans. The similarity in spectral shape of the 10-$\mu$m features extends to a similarity in both anhydrous silicate composition (olivine and pyroxene mixtures) and high regolith porosity. Asteroids that resemble the Trojans in this way include both D- \& P-types. Some major divergences from the spectra of the asteroids examined in this paper and previously published Trojan spectra are the lower continuum slopes and leftward skews of the 10 $\mu$m features seen in our sample. In the Trojans, the 10 $\mu$m feature tends to have a continuum slope that increases steeply with increasing wavelength and rightward skewing 10 $\mu$m features (\cite{martin-trojan}). The mineralogical significance of these spectral differences, however, is unclear. 

The distinctive spectrum of 368 Haidea is distinguished from the other asteroids in our sample and the Trojans (\cite{martin-trojan, emery_2006}) by the lack of evidence of spectral features associated with olivine, particularly the 11 $\mu$m peak. 368 Haidea is also notable for its low skew of 0.25. The shortward shift of 368 Haidea's features may be due to either a particularly pyroxene or phylosilicate rich composition, suggesting that 368 Haidea does not originate from the same source populations as the Trojans and other Main Belt D-types. Comparing 368 Haidea to the other D-types in our sample demonstrates the need for observations outside of the VNIR region, particularly for asteroids with few spectral features in the VNIR, to augment our understanding of the compositions of these asteroids. The recently launched JWST will be able to extend mid-infrared observations to a larger population of primitive asteroids, sampling cooler and fainter populations of primitive small Solar System bodies and supplement our understanding of ground-based VNIR asteroid taxonomies.

\section{Conclusions}

Combining new observations from the SOFIA observatory and archival data from the Spitzer Space Telescope, we examine the mid-infrared (5-40 $\mu$m) spectra of thirteen primitive Main Belt D- \& P- type asteroids. While these two asteroid populations resemble each other in the VNIR, a lack of strong spectral features in the VNIR limits the extent to which we can unambiguously determine their compositions. We analyze the spectra of these asteroids, including using techniques to characterize the 10-$\mu$m region developed in \cite{martin-trojan}, to compare the composition  and regolith properties of D- \& P-types in the Main Belt to the Jupiter Trojans. We find that the majority of D- \& P- type Main Belt asteroids in our sample resemble the Trojans in the mid-IR, with high regolith porosities and a silicate composition incorporating crystalline and amorphous olivines and pyroxenes, suggesting a common origin for primitive asteroids in the Main Belt and Trojan swarms. However, we also note some differences in the general shapes of the 10-$\mu$m features of Main Belt and Trojan asteroids including lower continuum slopes and leftward skewing 10 $\mu$m emission features among Main Belt asteroids as compared to Trojans. In addition, we identify a D-type asteroid, 368 Haidea, that is distinguished from both the Trojans and Main Belt asteroids in our sample by a spectrum lacking the 11 $\mu$m olivine-related emission feature. Despite their shared D-type classification, 368 Haidea does not resemble D-type Jupiter Trojans, suggesting that similarity in VNIR spectra does not guarantee a shared surface composition as reflected by silicate features in the mid-infrared.

\section{Acknowledgements}

The authors would like to thank Dr. Maggie McAdam for her assistance in preparing the SOFIA proposal. The authors would also like to thank the two anonymous reviewers whose input greatly improved the clarity of the manuscript. This work is based in part on observations made with the NASA/DLR Stratospheric Observatory for Infrared Astronomy (SOFIA). SOFIA is jointly operated by the Universities Space Research Association, Inc. (USRA), under NASA contract NNA17BF53C, and the Deutsches SOFIA Institut (DSI) under DLR contract 50 OK 2002 to the University of Stuttgart. Financial support for this work was provided by NASA through award \#08-0105 issued by USRA. This work is based in part on archival data obtained with the Spitzer Space Telescope, which was operated by the Jet Propulsion Laboratory, California Institute of Technology under a contract with NASA. This research has made use of data and/or services provided by the International Astronomical Union's Minor Planet Center. This majority of this work was completed at Northern Arizona University in Flagstaff, Arizona. Northern Arizona University sits at the base of the San Francisco Peaks, on homelands sacred to Native Americans throughout the region. We honor their past, present, and future generations, who have lived here for millennia and will forever call this place home.

\bibliography{sample631}{}
\bibliographystyle{aasjournal}

\end{document}